\documentclass[12pt,reqno]{amsart}

\usepackage{a4}
\usepackage[german,english]{babel}

\usepackage{amsmath, amssymb}
\usepackage{graphicx}                      
\newcommand{\color}[2][{}]{}        

\usepackage{bbm}         

\theoremstyle{plain}            
\newtheorem{theorem}{Theorem}[section]
\newtheorem*{theorem*}{Theorem}

\newtheorem*{maintheorem*}{Main Theorem}

\newtheorem{lemma}[theorem]{Lemma}
\newtheorem{corollary}[theorem]{Corollary}

\theoremstyle{definition}       
\newtheorem{definition}[theorem]{Definition}
\newtheorem{example}[theorem]{Example}

\theoremstyle{remark}           
\newtheorem{remark}[theorem]{Remark}

\newtheorem*{notation*}{Notation}

\newcommand{\Sec}[1]{Section~\ref{sec:#1}}

\newcommand{\App}[1]{Appendix~\ref{sec:#1}}

\newcommand{\Ex}[1]{Example~\ref{ex:#1}}
\newcommand{\Eq}[1]{Equation~\eqref{eq:#1}}
\newcommand{\Thm}[1]{Theorem~\ref{thm:#1}}
\newcommand{\Thms}[2]{Thms.~\ref{thm:#1},~\ref{thm:#2}}
\newcommand{\Lem}[1]{Lemma~\ref{lem:#1}}
\newcommand{\Cor}[1]{Corollary~\ref{cor:#1}}
\newcommand{\Rem}[1]{Remark~\ref{rem:#1}}
\newcommand{\Def}[1]{Definition~\ref{def:#1}}

\newcommand{\Fig}[1]{Figure~\ref{fig:#1}}

\numberwithin{equation}{section}

\DeclareMathOperator{\dist}   {dist}
\DeclareMathOperator{\dom}    {dom}

\DeclareMathOperator{\supp}   {supp}
\DeclareMathOperator{\vol}    {vol}

\newcommand{\spec}[2][{}]   {\sigma_{\mathrm{#1}}(#2)}
\newcommand{\essspec}[1]{\spec[ess] {#1}}
\newcommand{\disspec}[1]{\spec[disc]{#1}}

\newlength{\maxbreite}%
\newlength{\maxhoehe}%
\newlength{\maxtiefe}%

\newcommand{\stelldrueber}[3][0pt]{
  \settowidth{\maxbreite}{#3}%
  \settoheight{\maxhoehe}{#3}%
  \settodepth{\maxtiefe}{#2}%
  \addtolength{\maxhoehe}{\maxtiefe}%
  {\makebox[\maxbreite]{\raisebox{\maxhoehe}{\hspace{#1}#2}}%
  \makebox[0pt][r]{#3}}%
}

\newcommand{\overcirc}[1]       
{\stelldrueber[.45ex]{$\scriptscriptstyle\circ$}{${#1}$}}

\newcommand{\R}{\mathbb{R}} 
\newcommand{\C}{\mathbb{C}} 
\newcommand{\N}{\mathbb{N}} 
\newcommand{\Z}{\mathbb{Z}} 
\newcommand{\Sphere}{\mathbb{S}} 

\newcommand{\eps}{\varepsilon} 
\renewcommand{\phi}{\varphi}   
\newcommand{\e}{\mathrm e}  
\newcommand{\dd}{\,\mathrm d} 

\newcommand{\wt}{\widetilde}           
\newcommand {\qf}[1]{\mathfrak{#1}}    

\newcommand{\HS}{\mathcal H}           

\newcommand{\Sobsymb} {\mathsf H}      
\newcommand{\Contsymb} {\mathsf C}     
\newcommand{\Lsymb}    {\mathsf L}     
\newcommand{\lsymb}    {\ell}          

\newcommand{\Cont}[2][{}]{\Contsymb^{#1}({#2})}

\newcommand{\Lsqr}[2][{}]{\Lsymb_2^{#1}({#2})} 
 
\newcommand{\lsqr}[1]{\lsymb_2({#1})}           

\newcommand{\Sob}[2][1]{\Sobsymb^{#1}({#2})} 

\newcommand{\norm}[2][{}]{\|{#2}\|_{{#1}}}    
\newcommand{\normsqr}[2][{}]{\|{#2}\|^2_{#1}} 
\newcommand{\Bignorm}[2][{}]{\Bigl\|{#2}\Bigr\|_{#1}}     

\newcommand{\iprod}[3][{}]{\langle{#2},{#3}\rangle_{#1}}  

\newcommand{\set}[2]{\{ \, #1 \, | \, #2 \, \} } 

\newcommand{\Bigset}[2]{\Bigl\{ \, #1 \, \Bigl|\Bigr. \, #2 \, \Bigr\} }

\newcommand{\map}[3]{{#1}\colon{#2}\longrightarrow{#3}} 

\newcommand{\bd}  {\partial}                
\newcommand{\clo}[1]{\overline{{#1}}} 

\newcommand{\dcup}{\mathrel{\uplus}}               
\newcommand{\bigdcup}{\operatorname*{\biguplus}}

\newcommand{\restr}[1]{{\restriction}_{#1}} 

\newcommand{\conj}[1]{\overline {{#1}}}       

\newcommand{\1}{\mathbbm 1}                    
\newcommand{\Neu}{{\mathrm N}}              
\newcommand{\Dir}{{\mathrm D}}              
\newcommand{\laplacian}[2][{}]{\Delta_{{#2}}^{{#1}}} 

\newcommand{\EW}[3][{}]{\lambda^{{#1}}_{#2}({#3})}
\newcommand{\EWD}[2]{\EW[\Dir]{#1}{#2}}      
\newcommand{\EWN}[2]{\EW[\Neu]{#1}{#2}}      

\newcommand{\vxeps}{{\eps,v}}
\newcommand{\edeps}{{\eps,e}}

\begin{document}
\title[Spectral convergence of quasi-one-dimensional
  spaces]{Spectral convergence of non-compact quasi-one-dimensional
  spaces}

\author{Olaf Post} 
\address{Department of Mathematics, 
  University of Kentucky,
  751~Patterson~Office~Tower, 
  Lexington, Kentucky 40506-0027, USA}
\email{post@ms.uky.edu}
\date{\today}




\begin{abstract}
  We consider a family of non-compact manifolds $X_\eps$ (``graph-like
  manifolds'') approaching a metric graph $X_0$ and establish
  convergence results of the related natural operators, namely the
  (Neumann) Laplacian $\laplacian {X_\eps}$ and the generalised
  Neumann (Kirchhoff) Laplacian $\laplacian {X_0}$ on the metric
  graph.  In particular, we show the norm convergence of the
  resolvents, spectral projections and eigenfunctions. As a
  consequence, the essential and the discrete spectrum converge as
  well. Neither the manifolds nor the metric graph need to be compact,
  we only need some natural uniformity assumptions.  We provide
  examples of manifolds having spectral gaps in the essential
  spectrum, discrete eigenvalues in the gaps or even manifolds
  approaching a fractal spectrum. The convergence results will be
  given in a completely abstract setting dealing with operators acting
  in different spaces, applicable also in other geometric situations.
\end{abstract}

\maketitle


\section{Introduction}
\label{sec:intro}

\begin{figure}[h]
  \begin{center}
\begin{picture}(0,0)%
 \includegraphics{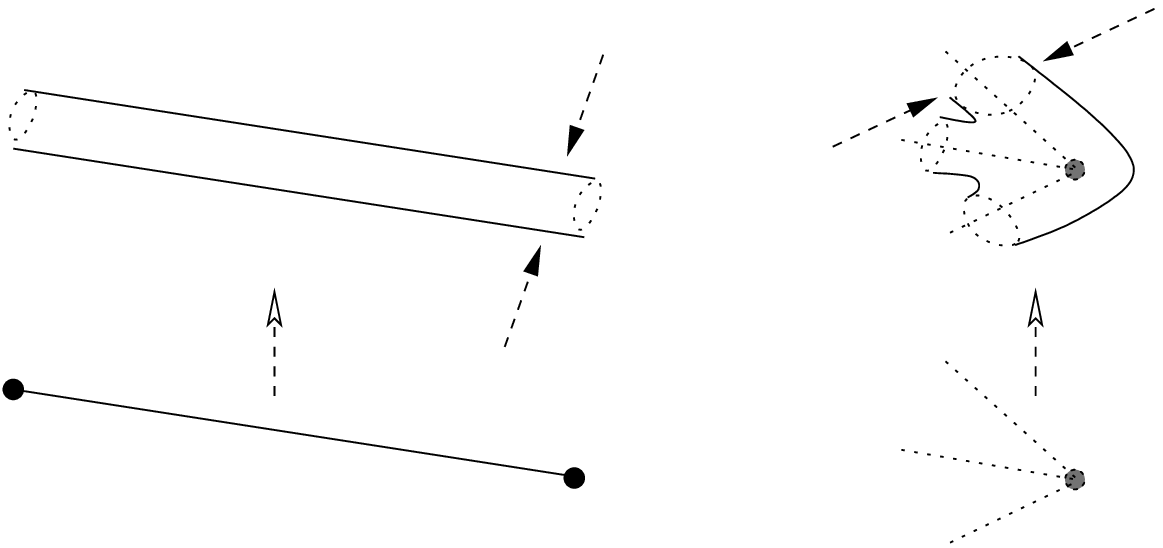}%
\end{picture}%
\setlength{\unitlength}{4144sp}%
\begin{picture}(5282,2500)(374,-1970)
  \put(1591,-586){$U_\edeps$}%
  \put(1561,-1674){$e$}%
  \put(5641,-331){$U_\vxeps$}%
  \put(5416,-1771){$v$}%
  \put(4579,374){$\eps$}%
  \put(3169,-436){$\eps$}%
\end{picture}
    \caption{The associated edge and vertex neighbourhoods with
      $F_\eps=\Sphere^1_\eps$, i.e., $U_\edeps$ and $U_\vxeps$ are
      $2$-dimensional manifolds with boundary.}
    \label{fig:edge.vertex}
  \end{center}
\end{figure}

\begin{figure}[htb]
  \begin{center}
    \begin{picture}(0,0)%
      \includegraphics{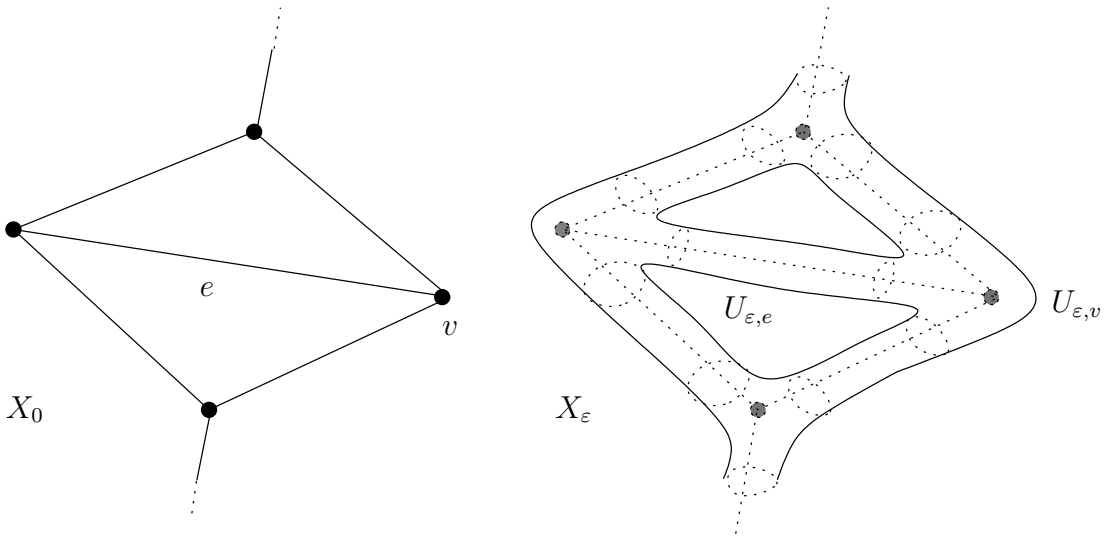}%
    \end{picture}%
    \setlength{\unitlength}{4144sp}%
    \begin{picture}(6263,3204)(308,-2913)
      \put(3601,-2176){$X_\eps$}
      \put(6571,-1546){$U_{\eps,v}$} 
      \put(4611,-1581){$U_{\eps,e}$}
      \put( 316,-2176){$X_0$}
      \put(2926,-1681){$v$}
      \put(1476,-1451){$e$}
    \end{picture}
    \caption{On the left, we have the graph $X_0$, on the right, the
      associated family of graph-like manifolds. Here,
      $F_\eps=\Sphere^1_\eps$ is the transversal section of radius
      $\eps$ and $X_\eps$ is a $2$-dimensional manifold.}
    \label{fig:mfd}
  \end{center}
\end{figure}

The aim of this article is to show that non-compact
quasi-one-dim\-ensional spaces can be approximated by the underlying
metric graph.  A \emph{metric} or \emph{quantum graph} is a graph
considered as one-dimensional space where each edge is assigned a
length. A \emph{quasi-one-dimensional space} consists of a family of
\emph{graph-like} manifolds, i.e., a family of manifolds $X_\eps$
shrinking to the underlying metric graph $X_0$. The family of
graph-like manifolds is constructed of building blocks $U_\vxeps$ and
$U_\edeps$ for each vertex $v \in V$ and edge $e \in E$ of the graph,
respectively (cf.~\Fig{edge.vertex}).
The cross section of the edge neighbourhood $U_\edeps$ as well as the
boundary component of $U_\vxeps$, where $U_\edeps$ meet, consists of a
manifold $F_\eps$ with radius of order $\eps$.  The cross section
could have a boundary resulting in a manifold $X_\eps$ \emph{with}
boundary. In addition, the vertex neighbourhoods $U_\vxeps$ are
assumed to be small.  The simplest example is the $\eps$-neighbourhood
$X_\eps$ of a quantum graph $X_0$ embedded in $\R^2$. In this case,
the cross section is $F_\eps=(-\eps,\eps)$.  A simple boundaryless
example is given by the \emph{surface} of a pipeline network according
to the underlying graph $X_0$ (cf.~\Fig{mfd}). Here, the cross section
consists of a circle of radius $\eps$.

On the graph-like manifold $X_\eps$ we consider the Laplacian $\wt
H:=\laplacian {X_\eps} \ge 0$ acting in the Hilbert space $\wt \HS :=
\Lsqr {X_\eps}$.  If $X_\eps$ has a boundary, we impose \emph{Neumann}
boundary conditions.  On the graph, we choose the natural Laplacian
$H:=\laplacian {X_0} \ge 0$, namely, the generalised Neumann
(Kirchhoff) Laplacian acting on each edge as a one-dimensional
weighted Laplacian (cf.\ Eq.~\eqref{eq:formaledge}). On each vertex,
we assume continuity and current conservation (cf.\ 
Eq.~\eqref{eq:kirchhoff}).  Note that $\laplacian {X_0}$ acts on
$\HS:=\oplus_e \Lsqr e$ where each edge $e$ is identified with the
interval $(0,\ell_e)$ ($0<\ell_e \le \infty$) --- in contrast to the
\emph{discrete} graph Laplacian acting as difference operator on the
space of vertices, $\lsqr V$. For a relation between these two
operators see~\Sec{disc.graph}.

In this article, we concentrate on the spectrum of such systems. Our
main result is the following:
\begin{maintheorem*}[\Thm{graph}]
  Suppose $X_\eps$ is a family of (non-compact) graph-like manifolds
  associated to a metric graph $X_0$. If $X_\eps$ and $X_0$ satisfy
  some natural uniformity conditions, then the resolvent of
  $\laplacian {X_\eps}$ converges in norm to the resolvent of
  $\laplacian {X_0}$ (with suitable identification operators) as $\eps
  \to 0$. In particular, the corresponding essential and discrete
  spectra converge uniformly in any bounded interval. Furthermore, the
  eigenfunctions converge as well.
\end{maintheorem*}

The \emph{uniformity conditions} are precisely stated in
\Sec{ass.graph}.  For example we need a global lower bound on the edge
length $\ell_e\ge \ell_0$ and a global upper bound on the vertex
degree $\deg v \le d_0$.  In the case when the graph $X_0$ is embedded
in $\R^2$ (cf.~\Sec{emb.graph}) the uniformity conditions mean in
particular, that we need a global bound on the curvature of an edge
and a global lower bound on the angle between two different edges at a
vertex \emph{although} both quantities do not enter into the limit
operator and space.

In contrast to previous articles (cf.~\cite{rubinstein-schatzman:01,
  kuchment-zeng:01, kuchment-zeng:03, exner-post:05}) we allow here
\emph{infinite} structures, i.e., we drop the condition of
\emph{compactness} of $X_\eps$ and $X_0$. Therefore, we cannot use the
variational principle in order to characterise the discrete spectrum.
The appropriate substitute is an abstract convergence criterion
provided in \App{abstr.mod}. The basic idea is to define a
``distance'' between the operators $\laplacian {X_0}$ and $\laplacian
{X_\eps}$ with suitable identification operators
(cf.~\Def{closeness}). We have formulated the abstract results fully
in terms of this ``distance'' in order to trace the parameter
dependence on $\eps$ of the operator $\laplacian {X_\eps}$, the
Hilbert space $\Lsqr {X_\eps}$ \emph{and} the identification operators
between the graph and the manifold. The ``distance'' can be calculated
in terms of the associated sesquilinear forms which makes the
verification quite simple in our main model. As a consequence, we show
norm resolvent convergence which implies all other convergence results
like convergence of the spectral projections, convergence of the
eigenfunctions and convergence of the spectra. Note that
in~\cite{rubinstein-schatzman:01, kuchment-zeng:01, kuchment-zeng:03,
  exner-post:05}, only convergence of eigenvalues has been
established. Our results here show, that the eigenvectors converge as
well.  We will show in a forthcoming paper that this abstract
convergence criterium has applications in other geometrical
situations.

A related result on \emph{non-compact} spaces has been established
in~\cite{saito:00}. Saito considered metric trees (allowing also
arbitrary small edges, i.e., no lower bound on $\ell_e$) together with
a suitable $\eps$-neighbourhood, but showed only \emph{weak}
convergence of the resolvents.  In~\cite{evans-saito:00} the authors
prove \emph{exact} relations (equality, inclusion) of the essential
spectrum of the Neumann Laplacian on a thickened tree (for fixed
$\eps$ in our notation) and the corresponding metric graph.

Our spectral convergence result has many applications in different
situations: First, we can consider graph-like manifolds as a kind of
toolbox in order to construct manifolds with prescribed spectrum, at
least approximately. For example, we are able to construct manifolds
with gaps in the essential spectrum (cf.~\Thms{gaps}{ew.graph}~) also
in the non-periodic case: Using the recent result on \emph{graph
  decoration} one can construct metric graphs with spectral gaps
(cf.~\cite{aizenman-schenker:00, beg:03, bgl:05, kuchment:05} and
\Sec{examples}). Our spectral convergence result then immediately
states that an associated graph-like manifold also has gaps. In the
periodic case (i.e., on covering spaces with compact quotient), we
have of course the same result once we ensure the existence of gaps on
the quantum graph. For the existence of spectral gaps on periodic
manifolds (not necessarily graph-like in our sense) we refer
to~\cite{post:03a, lledo-post:pre04, lledo-post:pre05} and the
references therein. The periodic case can often be reduced to the
spectral convergence on a compact space.

An example with arbitrary many gaps in a \emph{compact} spectral
interval is given by a fractal-like manifold in~\Thm{sierpinski}.  The
graph-like manifold is constructed according to a Sierpi\'nski graph.
It was shown in~\cite{teplyaev:98} that the discrete Laplacian on a
Sierpi\'nski graph has pure point spectrum which is purely essential
and of fractal nature. Using a nice relation between the spectrum of
the discrete graph Laplacian and the metric graph Laplacian with
constant edge length $\ell_e=\ell$ developed in~\cite{cattaneo:97}
(cf.~\Thm{disc.met}) we are able to construct a family of graph-like
manifolds $X_\eps$ such that the spectrum of $\laplacian {X_\eps}$
approaches a fractal set. In particular, the spectrum of $\laplacian
{X_\eps}$ has an arbitrary (a priori finite) number of spectral gaps
in the \emph{compact} interval $[0, \Lambda]$ provided $\eps$ is small
enough. Such fractal manifolds have been constructed in~\cite{bcg:01}
in order to provide examples of smooth spaces sharing properties of
fractal spaces in large scales (e.g.~heat kernel estimates).

Finally, our result shows rigorously, that the physically intuition of
modeling quasi-one-dimensional spaces by its singular limit is
correct, also on \emph{infinite} structures. Graph models have a long
history in modeling properties of networks, complicated organic
molecules~\cite{ruedenberg-scherr:53} or quite recently,
nanostructures, i.e., structures, which are too small to be considered
classically, but still too large to be described on a conventional
quantum level, see e.g.~\cite{aghh:05, kostrykin-schrader:99,
  kuchment:02, kuchment:04, kuchment:05}. On the one side quantum
graphs provide a solvable model in quantum mechanics in the sense that
many quantities can be calculated explicitely essentially by solving
systems of ODEs.  On the other side, the structure of a metric graph
is still rich enough in order to provide a good model for branched
structures.  For example, a spectral gap corresponds to ``forbidden
modes'', i.e., a particle with an energy in the gap cannot propagate
through the system. In this sense, transport properties on $X_\eps$
are approximately described by the quantum graph $X_0$.  Furthermore,
a bound state (of finite degeneracy) on $X_\eps$ (i.e., an
eigenfunction corresponding to a discrete eigenvalue) can be
approximated by its (mostly explicitly known) eigenvector on $X_0$
(cf.~\Thm{ew.graph}). In a forthcoming paper~\cite{exner-post:pre06}
we will deal with the convergence of resonances, i.e., eigenvalues of
a suitable dilated Hamiltonian. The methods needed there differ from
the ones given in~\App{abstr.mod} since the dilated operators are no
longer self-adjoint (even not normal).

With our methods here, we consider the discrete and the essential
spectrum only since they can be characterised by the dimension of
spectral projections. A finer analysis of the spectrum needs more
elaborated methods, such as scattering theory. Our results presented
here are considered as a first step in dealing with the
above-mentioned structures. We will concentrate on the relation
between scattering and transport properties on the two systems in a
forthcoming paper.

The paper is organised as follows.  In \Sec{graph} we define properly
graph-like manifolds and metric graphs and show that the abstract
convergence result can be used in this situation for
$(H,\HS)=(\laplacian{X_0}, \Lsqr{X_0})$ and $(\wt H,\wt \HS) =
(\laplacian{X_\eps},\Lsqr{X_\eps})$.  In \Sec{ex.app} we discuss
various examples of graph-like manifolds to which our result applies.
We also derive several consequences of the spectral convergence.  In
\App{abstr.mod} we develop the abstract framework in order to show the
spectral convergence for arbitrary pairs $(H,\HS)$ and $(\wt H, \wt
\HS)$ being at a ``distance'' $\delta$ to each other.

\section{Graph-like manifolds}
\label{sec:graph}

In this section we apply the abstract setting developed in
\App{abstr.mod} to the example of a family of manifolds $X_\eps$
converging to a (metric) graph $X_0$.  This situation has already been
treated in a quite general way in~\cite{exner-post:05} based
on~\cite{kuchment-zeng:01, rubinstein-schatzman:01} with the only
restriction that the graph is \emph{compact} (i.e., finite and each
edge has finite length) and each manifold $X_\eps$ is \emph{compact}.
Under these assumptions, the spectra of the operators considered are
purely discrete (for a precise definition see below). The main result
in~\cite{exner-post:05} states that the $k$-th eigenvalue of the
Laplacian $\laplacian {X_\eps}$ converges to the $k$-th eigenvalue of
the limit operator. The proof uses the min-max principle and
comparison of the appropriate Rayleigh quotients.

If the manifold and the metric graph are non-compact, more elaborated
methods are needed. Namely, we establish in \App{abstr.mod} norm
resolvent convergence from which all other convergence results follow.
The norm resolvent convergence is reduced to the verification of
several natural conditions provided in \Def{closeness}. In order that
these conditions are satisfied we need the uniformity
assumptions~\eqref{eq:deg.bd}--\eqref{eq:vol.vertex} in our model. The
eigenvalue convergence already proven in~\cite{exner-post:05} appears
as a special case (cf.\ \Cor{ev.conv}).

\subsection{Metric graphs}
\label{sec:met.graph}

Let us first describe the metric graph $X_0$ and the family of
graph-like manifolds $X_\eps$; the necessary assumptions in order that
the convergence results hold will be given later.  Suppose
$X_0=(V,E,\bd, \ell)$ is a countable, connected metric graph, i.e.,
$V$ denotes the set of vertices, $E$ the set of edges and $\map \bd E
{V \times V}$, $\bd e = (\bd_+e,\bd_-e)$ denotes the pair of the end
point and the starting point of the edge $e$.  For each vertex $v \in
V$ we denote by
\begin{equation*}
  E_v^\pm := \set {e \in E} {\bd_\pm e = v}
\end{equation*}
the edges starting ($-$)/ending ($+$) at $v$. Let $E_v := E_v^+ \dcup
E_v^-$ be the \emph{disjoint} union of all edges emanating at $v$.
The \emph{degree} of a vertex $v$ is the number of vertices emanating
from $v$, i.e.,
\begin{equation}
  \label{eq:degree}
  \deg v := |E_v| = |E_v^+| + |E_v^-|.
\end{equation}
We assume that $X_0$ is \emph{locally finite}, i.e., $\deg v \in \N$.
Note that we allow loops, i.e. edges $e$ with $\bd_+ e = \bd_- e = v$.
A loop $e$ will be counted twice in $\deg v$ and occurs twice in $E_v$
due to the disjoint union. In addition, we assume that $\bd e$ always
consists of two elements, even if $\bd_-e = \bd_+e = v$ for a loop
$e$.  We also allow multiple edges, i.e., edges $e_1 \ne e_2$ having
the same starting and end points.

Finally, $\map \ell E {(0,\infty]}$ assigns a length $\ell_e$ to each
edge $e \in E$, making the graph $(V,E,\bd)$ a \emph{metric} or
\emph{quantum} graph. Clearly, $X_0$ becomes a metric space.  We
identify each edge $e$ with the interval $(0,\ell_e)$.  In the case of
an infinite edge, a ``lead'', (i.e., $\ell_e=\infty$) we assume that
there is only one vertex $\bd e = \bd_-e$ at the end corresponding to
$0$, i.e., there is no vertex at $\infty$.  For a general survey on
quantum graphs consult e.g.~\cite{kuchment:04, kuchment:05}. We stress
that our graphs need by no way to be embedded in some Euclidean space.
\begin{remark}
  Note that for a metric graph, the notion ``compact'' and ``finite''
  have a different meaning: A \emph{finite} metric graph is a graph
  with finitely many vertices and edges, whereas a \emph{compact}
  graph must in addition have finite edge length for each edge.
  Therefore, a compact metric graph is finite but not vice versa
  (think e.g.\ of a star-shaped metric graph with one vertex and a
  finite number of leads attached to the vertex).
\end{remark}
We also assign a \emph{density} $p_e$ to each edge $e \in E$, i.e., a
measurable function $\map {p_e} e {(0,\infty)}$.  For simplicity, we
assume that $p_e$ is smooth in order to obtain a smooth metric in the
graph-like manifold. The data $(V,E,\bd,\ell,p)$, $p=(p_e)_e$ describe
a \emph{weighted} metric graph.

The Hilbert space associated to such a graph is
\begin{equation*}
  \HS := \Lsqr {X_0} = 
  \bigoplus_{e \in E} \Lsqr e
\end{equation*}
which consists of all functions $f$ with finite norm
\begin{equation*}
  \normsqr f = \normsqr[X_0] f = 
  \sum_{e \in E} \normsqr[e] {f_e} =
  \sum_{e \in E} \int_e |f_e(x)|^2 p_e(x) \dd x.
\end{equation*}
We define the limit operator $H$ via the quadratic form
\begin{displaymath}
  \qf h(f) := 
  \sum_{e \in E} \normsqr[e] {f'_e} =
  \sum_{e \in E} \int_e |f'_e(x)|^2 p_e(x) \dd x
\end{displaymath}
for functions $f$ in
\begin{equation*}
  \HS_1 := \Sob {X_0} :=
  \Cont {X_0} \cap \bigoplus_{e \in E} \Sob e.
\end{equation*}
Note that a weakly differentiable function on an interval $e$, i.e.,
$f_e \in \Sob e$, is automatically continuous. Therefore, the
continuity is only a condition at each vertex. Furthermore, $\qf h$ is
a closed form, i.e., $\HS_1$ together with the norm
\begin{equation*}
  \normsqr[1] f = \normsqr[1, X_0] f := \normsqr[X_0] f + \qf h(f)
\end{equation*}
is complete.

The associated self-adjoint, non-negative operator $H=\laplacian
{X_0}$ is given by
\begin{equation}
\label{eq:formaledge}
  (\laplacian {X_0} f)_e = -\frac 1{p_e}\big(p_e \, f_e'\big)'
\end{equation}
on each edge $e$.  If we assume the global lower
bound~\eqref{eq:length} (cf.\ page~\pageref{eq:length}) on the length
$\ell_e$ of the edge $e$ then the domain $\HS_2$ of $H=\laplacian
{X_0}$ consists of all functions $f \in \Lsqr {X_0}$ such that
$\laplacian {X_0} f \in \Lsqr {X_0}$
(cf.~e.g.~\cite[Thm.~17]{kuchment:04}). Furthermore, each function $f$
satisfies the so-called \emph{(generalised) Neumann boundary
  condition} (sometimes also named \emph{Kirchhoff}) at each vertex
$v$, i.e., $f$ is continuous at $v$ and
\begin{equation}
  \label{eq:kirchhoff}
  \sum_{e \in E_v} p_e(v) f'_e(v) = 0
\end{equation}
for all vertices $v \in V$ where the derivative is taken \emph{away}
from the vertex, i.e.\ we set $f'_e(v) := f'_e(0)$ if $v=\bd_-e$ and
$f'_e(v) := -f'_e(\ell_e)$ if $v=\bd_+e$ (considering $f_e$ as
function on the interval $(0,\ell_e)$).  We will call $\laplacian
{X_0}$ the \emph{(generalised) weighted Neumann Laplacian} on $X_0$.
For details on operators on non-compact or infinite metric graphs we
refer e.g.  to~\cite{kuchment:04, kuchment:05}.

\subsection{Graph-like manifolds}
\label{sec:graph.mfd}

Let $X_0$ be a weighted metric graph as defined in the previous
section.  The corresponding family of graph-like manifolds $X_\eps$ is
given as follows: For each $0 < \eps \le \eps_0$ we associate with the
graph $X_0$ a connected Riemannian manifold $X_\eps$ of dimension $d
\ge 2$ with or without boundary equipped with a metric $g_\eps$ to be
specified below\footnote{The boundary of $X_\eps$ (if there is any)
  need not to be smooth; we allow singularities on the boundary of the
  vertex neighbourhood $U_\vxeps$, see e.g.\ \Sec{emb.graph}
  and~\Fig{graph.emb}}. We suppose that $X_\eps$ is the union of the
closure of open subsets $U_\edeps$ and $U_\vxeps$ such that the
$U_\edeps$ and $U_\vxeps$ are mutually disjoint for all possible
combinations of $e \in E$ and $v \in V$, i.e.,
\begin{equation}
  \label{eq:disjcup}
  X_\eps = \bigcup_{e \in E} \clo U_\edeps \cup 
   \bigcup_{v \in V} \clo U_\vxeps.
\end{equation}
We think of $U_\edeps$ as the thickened edge $e$ and of $U_\vxeps$ as
the thickened vertex $v$ (see Figures~\ref{fig:edge.vertex}
and~\ref{fig:mfd}).  Note that Figure~\ref{fig:mfd} describes the
situation only roughly, since it assumes that $X_\eps$ is embedded in
$\R^\nu$. More correctly, we should think of $X_\eps$ as an abstract
manifold obtained by identifying the appropriate boundary parts of
$U_\edeps$ and $U_\vxeps$ via the connection rules of the graph $X_0$.
This manifold need not to be embedded, but the situation when $X_\eps$
is a submanifold of $\R^\nu$ ($\nu \ge d$) can be viewed also in this
abstract context; note that the $\eps$-neighbourhood of an embedded
metric graph in $\R^d$ is also included as example
(cf.~\cite{exner-post:05} and \Sec{emb.graph}).

As a matter of convenience we assume that $U_\edeps$ and
$U_\vxeps$ are independent of $\eps$ as manifolds, i.e., only their
metrics $g_\eps$ depend on $\eps$.  This can be achieved in the
following way: for the edge regions we assume that $U_\edeps$ is
diffeomorphic to $U_e := e \times F$ for all $0 < \eps \le \eps_0$
where $F$ denotes a compact and connected manifold (with or without a
boundary) of dimension $m:=d-1$.  We fix a metric $h$ on $F$ and
assume for simplicity that $\vol F = 1$.

For the vertex regions we assume that the manifold $U_\vxeps$ is
diffeomorphic to an $\eps$-independent manifold $U_v$ for $0 < \eps
\le \eps_0$.  Pulling back the metric to the diffeomorphic manifold
$U_e$ resp.\ $U_v$ we may assume that the underlying differentiable
manifold is independent of $\eps$.  Therefore, $U_\edeps \cong (U_e,
g_\edeps)$ and $U_\vxeps = (U_v, g_\vxeps)$.

We use the obvious notation for functions $u$ on $X_\eps$ like
$u_e$ and $u_v$ as restrictions on $U_\edeps$ and
$U_\vxeps$, respectively.  The corresponding Hilbert space is then
\begin{equation*}
  \wt \HS := \Lsqr {X_\eps} = 
  \bigoplus_{e \in E} \Lsqr {U_\edeps} \oplus
      \bigoplus_{v \in V} \Lsqr {U_\vxeps}
\end{equation*}
which consists of all functions $u$ with finite norm
\begin{multline*}
  \normsqr u =
  \normsqr[X_\eps] u = 
  \sum_{e \in E} \normsqr[U_\edeps] {u_e} +
     \sum_{v \in V} \normsqr[U_\vxeps] {u_v} \\ =
  \sum_{e \in E} 
     \int_{e \times F} |u_e|^2 \det g_\edeps^{1/2} \dd x \dd y
   + \sum_{v \in V} 
        \int_{U_v} |u_v|^2 \det g_\vxeps^{1/2} \dd z
\end{multline*}
where $y$ and $z$ represent coordinates of $F$ and $U_v$,
respectively.

The operator $\wt H$ we are considering will be the Laplacian on
$X_\eps$, i.e., $\wt H = \laplacian {X_\eps}$.  If $F$ has non-trivial
boundary, we assume Neumann boundary conditions on the boundary part
coming from $\bd F$. We define $\laplacian {X_\eps}$ via its quadratic
form $\wt{\qf h}$ given by
\begin{equation}
  \label{eq:def.quad.mfd}
  \wt{\qf h} (u) = \int_{X_\eps} |\dd u|_{g_\eps}^2 \dd X_\eps
\end{equation}
for functions $u \in \wt \HS_1 = \Sob {X_\eps}$ where the latter space
denotes the completion of the space of smooth functions with
\emph{bounded} support w.r.t.\ the norm
\begin{equation}
  \label{eq:def.quad.norm}
  \normsqr[1] u = 
  \normsqr[1, X_\eps] u :=
  \normsqr[X_\eps] u + \wt{\qf h}(u).
\end{equation}

\subsection{Quasi-unitary operators}
\label{sec:quasi-unitary}

Let us fix the identification operators $\map J \HS {\wt \HS}$ and
$\map {J'} {\wt \HS} \HS$ and their analogues on the quadratic form
domains. Roughly speaking, $J$ extends the function $f$ at $x \in e$
constantly onto the cross section $F_\edeps(x):=(\{x\} \times
F,h_\edeps) \subset (U_e, g_\edeps)$, where $h_\edeps$ is the induced
metric of the restriction, and $J'$ is the transversal average of $u$
at $x$, i.e., the Fourier coefficient of $u(x,\cdot)$ w.r.t.\ the
first (constant) eigenfunction on $F_\edeps(x)$. We will first show
what estimates are necessary in order that $J$, $J'$ and their
quadratic form analogues become \emph{quasi-unitary} in the sense of
\Def{closeness}. In a second step we provide the necessary assumptions
on the graph (\Sec{ass.graph}) and on the manifold (\Sec{ass.mfd}).
Finally, we provide some necessary estimates (\Sec{est.graph}) and
finish the proof of quasi-unitarity.

We define the operator $\map J \HS {\wt \HS}$ by
\begin{equation}
  \label{eq:gr.trans.op}
  J f (z):=
  \begin{cases}
    \eps^{-m/2} f_e(x) & \text{if $z=(x,y) \in U_e$},\\
    0 & \text{if $z \in U_v$}
  \end{cases}
\end{equation}
and the operator $\map {J_1} {\HS_1} {\wt \HS_1}$ by
\begin{equation}
  \label{eq:gr.trans.op1}
  J_1 f (z):=
  \begin{cases}
    \eps^{-m/2} f_e(x) & \text{if $z=(x,y) \in U_e$},\\
    \eps^{-m/2} f(v)   & \text{if $z \in U_v$}
  \end{cases}
\end{equation}
Note that the latter operator is well-defined since functions in
$\HS_1$ are continuous (cf.\ \Lem{point.vx}).  For the operators in
the opposite direction, we first introduce the following averaging
operators
\begin{gather*}
  (N_e u)(x) :=
   \iprod[F]{\phi_{F,1}} {u_e(x,\cdot)} =
   \int_F u_e(x,y) \dd F(y),\\
   C_v u := 
   \iprod[U_v] {\phi_{U_v,1}} {u_v} =
   \frac 1 {\vol U_v} \int_{U_v} u \dd U_v
\end{gather*}
for $u \in \wt \HS = \Lsqr {X_\eps}$ giving the coefficient
corresponding to the first (transversal) eigenfunction $\phi_1$ on
$U_e$ resp.\ $U_v$. Note that these eigenfunctions are constant and
that $\vol F = 1$.

We define $\map {J'} {\wt \HS} \HS$ by
\begin{equation}
  \label{eq:gr.trans.op.}
  (J' u)_e (x):=
    \eps^{m/2} (N_e u)(x),
      \qquad x \in e
 \end{equation}
and the operator $\map {J_1'} {\wt \HS_1} {\HS_1}$ by
\begin{multline}
  \label{eq:gr.trans.op1.}
  (J_1' u)_e (x):=
    \eps^{m/2} \Bigl[
        N_e u (x) \\ + 
        \rho^+_e(x) \bigl[ C_{\bd_+ e} u - N_e u (\bd_+ e) \bigr] +
        \rho^-_e(x) \bigl[ C_{\bd_- e} u - N_e u (\bd_- e) \bigr]
    \Bigr]
\end{multline}
for $x \in e$.  Here, $\map {\rho^\pm_e} \R {[0,1]}$ are the
continuous, piecewise affine functions given by
\begin{equation}
  \label{eq:def.cut.off}
  \rho^+_e(\bd_+ e)=1           \quad \text{and} \quad
  \rho^+_e(x)=0
      \quad \text{for all $\dist(x, \bd_+ e) \ge \min \{1, \ell_e/2\}$}
\end{equation}
and similarly for $\rho^-_e$ and $\bd_- e$.  Note that $(J_1' u)_e(v)
= C_v u$ for $v = \bd_\pm e$. In particular, $J_1' u$ is a continuous
function on $X_0$.  Again, the operator $J_1'$ is only defined on $\wt
\HS_1 = \Sob {X_\eps}$.

The closeness assumptions of \Sec{ass} now reads as follows:
\begin{gather}
  \label{eq:j.scale.gr}
  \normsqr{Jf - J_1f} = 
  \sum_{v \in V} \eps^{-m} \vol U_\vxeps |f(v)|^2\\
  \label{eq:j.scale..gr}
  \tag{\ref{eq:j.scale.gr}'}
  \normsqr{J' u - J_1' u} =
  \sum_{e \in E} \sum_ {v \in \bd e} 
     \eps^m \normsqr[e]{\rho^\pm_e} \, | C_v u - N_e u(v)|^2 \\
  \label{eq:j.adj.gr}
  \begin{split}
     |\iprod {Jf} u &- \iprod f {J' u} | \\ {}=
      &\Bigl| \sum_{e \in E} \int\limits_{e \times F} 
              \conj f(x) u(x,y) \, \eps^{-m/2}
      \bigl[ \dd U_\edeps(x,y) - 
              \eps^m \dd F(y) \, p_e(x) \dd x \bigr] \Bigr|
     \end{split}
\end{gather}
\begin{gather}
  \label{eq:j.comm.gr}
  \begin{split}
     |\wt{\qf h} (J_1 f, u) - \qf h(f, J_1' & u)| \\ {}=
      \Bigl| \sum_{e \in E} \int\limits_{e \times F} 
             \conj f'(x) \, \partial_x &u(x,y) \, \eps^{-m/2}
      \bigl[ g_\edeps^{xx} \dd U_\edeps(x,y) - 
              \eps^m \dd F(y) \, p_e(x) \dd x \bigr] \\
      &\qquad\qquad{}-\sum_{e \in E} \sum_{v \in \bd e} 
             \eps^{-m/2} (C_v u - N_e u(v)) 
                 \iprod[e] {f_e'}{(\rho^\pm_e)'} \Bigr|
     \end{split}\\
  \label{eq:j.inv..gr}
  \normsqr{J J' u - u} =
  \sum_{e \in E} \normsqr[U_\edeps] {N_e u - u} +
    \sum_{v \in V} \normsqr[U_\vxeps] u\\
  \label{eq:j.bdd.gr}
  \normsqr{J f} = 
  \sum_{e \in E} \int_{e \times F} |f(x)|^2 
         \eps^{-m} \dd U_\edeps(x,y)\\
  \label{eq:j.bdd..gr}
  \tag{\ref{eq:j.bdd.gr}'}
  \normsqr{J' u} \le
  \int_{e \times F} |u(x,y)|^2 \, \eps^m \dd F(y) \, p_e(x) \dd x 
\end{gather}
Here, the sign in $\rho^\pm_e$ is used according to $v=\bd_\pm e$.
Note that $J' J f = f$. From the RHS we can deduce the necessary
assumptions given precisely in the next section. For example,
from~\eqref{eq:j.scale.gr} it follows that we must have $\vol
U_\vxeps=o(\eps^m)$, and from~\eqref{eq:j.adj.gr}
and~\eqref{eq:j.comm.gr} we see that $g_\edeps$ must be close to a
product metric on $U_e = e \times F$.

\subsection{Assumptions on the graph}
\label{sec:ass.graph}
We precise here the necessary assumptions in order to estimate the RHS
of~\eqref{eq:j.scale.gr}--\eqref{eq:j.bdd.gr}
and~\eqref{eq:j.bdd..gr}. For the graph data we require that the
degree is uniformly bounded, i.e., that there exists $d_0 \in \N$ such
that
\begin{equation}
  \label{eq:deg.bd}
  \tag{G1}
  \deg v \le d_0, \qquad v \in V.
\end{equation}
We assume in addition that there is a uniform lower bound on the set
of length, i.e., there exists $\ell_0>0$ (without loss of generality
$\ell_0 \le 1$) such that
\begin{equation}
  \label{eq:length}
  \tag{G2}
  \ell_e \ge \ell_0 \qquad \text{for all $e \in E$.}
\end{equation}
We assume that the density function $p_e$ is uniformly bounded, i.e.,
there exist constants $p_\pm > 0$ such that
\begin{equation}
  \label{eq:density}
  \tag{G3}
  \begin{split}
    p_- \le p_e(x), \qquad &
       \dist(x, \bd_\pm e) \le \min \{1,\ell_e/2\}, \qquad
        e \in E,\\
   p_e(x) \le p_+, \qquad &x \in e, \qquad e \in E.
  \end{split}
\end{equation}
Since $r_e(x):=p_e(x)^{1/m}$ will correspond to the radius of the
cross section of $U_\edeps$ at $x \in e$, we also denote $r_\pm :=
p_\pm^{1/m}$ the maximal/minimal radius. We want to stress that we
allow a sequence of edges $e_n$ such that $\ell_{e_n} \to \infty$ or
even external edges (i.e., edges with infinite length). In both cases,
we do not impose a global \emph{lower} bound on the density function
$p_e=r_e^m$. E.g., a horn-like shape of radius $r_e(x)=r_{e,0}
x^{-\beta}$ for the associated edge neighbourhood $U_\edeps$ is
allowed for an external edge $e$ (cf.\ also \Rem{horn}).

\begin{definition}
  \label{def:unif.graph}
  A \emph{uniform} weighted metric graph is a weighted metric graph
  $X_0=(V,E,\bd,\ell,p)$ satisfying
  \eqref{eq:deg.bd}--\eqref{eq:density}.
\end{definition}
From these assumptions we conclude the following estimates:

\begin{lemma}
  \label{lem:ed.vx}
  Suppose that $(a(v))_{v \in V}$ is a family of non-negative
  numbers.  Then
  \begin{equation}
  \label{eq:sum.ed.vx}
    \sum_{e \in E} \sum_{v \in \bd e} a(v) =
    \sum_{e \in E} \bigl(a(\bd_+ e) + a(\bd_- e)\bigr) =
    \sum_{v \in V} (\deg v) a(v) \le
    d_0 \sum_{v \in V} a(v)
  \end{equation}
  due to~\eqref{eq:deg.bd}. Furthermore,
  \begin{equation}
  \label{eq:sum.vx.ed}
    \sum_{v \in V} \sum_{e \in E_v} b_e = 2 \sum_{e \in E} b_e
  \end{equation}
  for a family $(b_e)_{e \in E}$.
\end{lemma}
\begin{proof}
  Inequality~\eqref{eq:sum.ed.vx} is clear, and from the disjoint
  union
  \begin{equation*}
    E = \bigdcup_{v \in V} E_v^+ = \bigdcup_{v \in V} E_v^-
  \end{equation*}
  the second equality follows immediately.
\end{proof}
The next lemma is needed in order to estimate $f(v)$:
\begin{lemma}
  \label{lem:point.vx}
  We have
  \begin{equation*}
    \sum_{v \in V} |f(v)|^2 \le \frac 4 {\ell_0 p_-} \normsqr[1] f
  \end{equation*}
  for all $f \in \HS_1 = \Sob {X_0}$.
\end{lemma}
\begin{proof}
  The estimate follows easily from
  \begin{equation}
    \label{eq:point.vx}
    |f(0)|^2 \le
    \frac 2 {p_-} \max \Bigl\{ \frac 1 \ell, \ell \Bigr\}
       \int_0^\ell \bigl( |f(x)|^2 + |f'(x)|^2 \bigr) p(x) \,
         \dd x
  \end{equation}
  where $f \in \Sob{0,\ell}$ and $p_- := \inf_{0 \le x \le \ell} p(x)$
  (cf.~\cite[Lemma~8]{kuchment:04}) applied to $\ell := \min \{ 1,
  \ell_e/2 \}$ together with~\eqref{eq:length} and~\eqref{eq:density}.
\end{proof}

Finally, we can estimate the cut-off function $\rho^\pm_e$
using~\eqref{eq:length} and~\eqref{eq:density}:
\begin{lemma}
  \label{lem:cut.off}
  The estimate
  \begin{equation}
  \label{eq:cut.off}
    \normsqr[e] {\rho^\pm_e} \le p_+ \qquad \text{and} \qquad
    \normsqr[e] {(\rho^\pm_e)'} \le \frac {2 p_+} {\ell_0}
  \end{equation}
  holds for all $e \in E$.
\end{lemma}

\subsection{Assumptions on the manifold}
\label{sec:ass.mfd}
Guided by the classical example of an embedded graph
(cf.~\Sec{emb.graph} and~\eqref{eq:met.emb}) we assume that the metric
$g_\edeps$ on the edge neighbourhood $U_e=e \times F$ is given as a
perturbation of the product metric
\begin{equation}
  \label{eq:def.met.edge}
  \overline g_\edeps := \dd x^2 + \eps^2 \, r_e^2(x) \, h(y), \qquad 
  (x,y) \in U_e = e \times F
\end{equation}
with
\begin{equation*}
  \label{eq:def.rad}
  r_e(x) := (p_e(x))^{1/m}
\end{equation*}
where $h$ is the fixed metric on $F$, $m=\dim F = d-1$ and $p_e$ is
the density function of the metric graph on the edge $e$.

We denote by $G_\edeps$ and $\overline G_\edeps$ the $d \times d$-matrices
associated to the metrics $g_\edeps$ and $\overline g_\edeps$ with respect
to the coordinates $(x,y)$ and assume that the two metrics coincide up
to an error term as $\eps\to 0$, more specifically
\begin{equation}
  \label{eq:asym.met.ed}
  \tag{G4}
  G_\edeps = \overline G_\edeps +
    \begin{pmatrix}
      o(1) & o(\eps)r_e \\ o(\eps)r_e &
      o(\eps^2)r_e^2
    \end{pmatrix} =
    \begin{pmatrix}
      1 + o(1) & o(\eps)r_e \\ o(\eps)r_e &
      (\eps^2   H + o(\eps^2))r_e^2
    \end{pmatrix}
\end{equation}
\emph{uniformly} on $U_e$. We also assume that these error estimates
are uniform in $e$, i.e., that $o(\eps^i)$ does not depend on the edge $e
\in E$. As in \cite[Lemma~4.3]{exner-post:05} (replacing $\eps$ by
$\eps r_e\le \eps r_+$) we can show the following estimates
\begin{gather}
  \label{eq:met.vol}
    \dd U_\edeps(x,y) = (1 + o_1(1)) \eps^m \dd F(y) \, p_e(x) \dd x \\
  \label{eq:met.1st.comp}
    g_\edeps^{xx}  \!:= (G_\edeps^{-1})_{xx} = 1 + o_2(1) \\
  \label{eq:met.1st.der}
    |\dd_x u|^2 \le O_3(1) |\dd u|_{g_\edeps}^2\\
  \label{eq:met.2nd.der}
    |\dd_F u|_h^2 \le o_4(\eps) |\dd u|_{g_\edeps}^2
\end{gather}
where $\dd_x$ and $\dd_F$ are the (exterior) derivatives with respect
to $x \in e$ and $y \in F$, respectively. Here, $o_1(1)$ and $o_2(1)$
depend only on $o(\eps^j)$ in~\eqref{eq:asym.met.ed} whereas $O_3(1)$
and $o_4(1)$ depend also on $r_+$.  The index $i$ in $o_i(\cdot)$ is
added in order to trace the error estimates in the formulas below. All
the estimates are uniform on $U_e$ and uniform in $e \in E$ as $\eps
\to 0$.

On the vertex neighbourhood $U_v$ we assume that the metric $g_\vxeps$
satisfies
\begin{equation}
  \label{eq:met.vx}
  \tag{G5}
  c_- \eps^2 g_v \le g_\vxeps \le c_+ \eps^{2\alpha} g_v
\end{equation}
in the sense that there are constants $c_-, c_+>0$ \emph{independent}
of $v$ such that
\begin{displaymath}
  c_- \eps^2 g_v(z)(w,w) \le 
  g_\vxeps(z)(w,w) \le 
  c_+ \eps^{2\alpha} g_v(z)(w,w)
\end{displaymath}
for all $w \in T_z U_v$ and all $z \in U_v$ where $g_v$ is the metric
$g_\vxeps$ with $\eps=1$. The number $\alpha$ in the exponent is
assumed to satisfy the inequalities
 \begin{equation}
  \label{eq:est.alpha}
  \tag{G6}
  \frac {d-1} d < \alpha \le 1.
\end{equation}
In addition, we assume that
\begin{equation}
  \label{eq:vol.vertex}
  \tag{G7}
  c_{\vol} := \sup_{v \in V} \vol U_v < \infty 
    \quad \text{and} \quad
  \lambda_2 := \inf \EWN 2 {U_v} > 0
\end{equation}
where $\EWN 2 {U_v}$ denotes the second (i.e., the first non-zero)
Neumann eigenvalue of $\laplacian {U_v}$.

\begin{definition}
  \label{def:unif.mfd}
  A family of graph-like manifolds $X_\eps$ w.r.t.\ the uniform metric
  graph $X_0$ will be called \emph{uniform}
  if~\eqref{eq:asym.met.ed}--\eqref{eq:vol.vertex} are satisfied.
\end{definition}
We will discuss several examples of uniform metric graphs and
graph-like manifolds in \Sec{ex.app}. Let us just finish this
subsection with a few comments on the assumptions.

\begin{remark}
  \label{rem:unif.mfd}
  \begin{enumerate}
  \item The condition~\eqref{eq:asym.met.ed} is motivated by the
    classical example of a curved edge embedded in $\R^2$,
    cf.~\Sec{emb.graph} and~\eqref{eq:met.emb}.
  \item We have assumed an upper bound of the density $p_e$ (i.e., the
    radius function $r_e$) on the \emph{whole} edge
    in~\eqref{eq:density}. A careful analysis of the proof
    of~\eqref{eq:met.vol}--\eqref{eq:met.2nd.der} shows that the first
    two estimates remain globally true even if $p_e$ has no global
    bound $p_+$ on $e$. But the last two estimates need the global
    bound $p_e(x) \le p_+$ for all $x \in e$ and $e \in E$. Therefore,
    an infinite edge cannot have a neighbourhood $U_\edeps$ with
    growing radius $r_e(x) \to \infty$ such as a conical end. It is
    also forbidden that a sequence of edges $e_n$ has neighbourhoods
    with radius functions $r_{e_n}$ unbounded in $n$.
  \item 
    \label{rem:bottle.neck}
    Note that the upper estimate in assumption~\eqref{eq:met.vx} does
    not apply to points \emph{at the border} of $\bd U_v$, since we
    still assume that $g_\vxeps$ is the restriction of a \emph{global}
    metric $g_\eps$ on $X_\eps$; and the cross-section on $U_\edeps$
    has a metric of order $O(\eps^2) h$.  Nevertheless it is not
    excluded that $g_\vxeps$ scales differently \emph{away} from $\bd
    U_v$ (for a detailed discussion of such scalings
    cf.~\cite[Sec.~6]{exner-post:05}).
  \item
    \label{rem:critical}
    The reason for the critical exponent $(d-1)/d$
    in~\eqref{eq:est.alpha} is roughly the following: If $\alpha$
    satisfies~\eqref{eq:est.alpha} then $\vol U_\vxeps \le
    O(\eps^{\alpha d})$ decays faster than $\vol U_\edeps =
    O(\eps^{d-1})$. Other decay rates are discussed
    in~\cite{kuchment-zeng:03, exner-post:05}.
  \item Assumption~\eqref{eq:vol.vertex} roughly assures that $U_v$
    remains small (as family in $v \in V$): Suppose that there is an
    infinite sequence $(v_n) \subset V$ such that $U_{v_n}= U_{v_0}$
    as sets and that $g_{v_n} = \rho_n^2 g$ for a sequence $\rho_n \to
    \infty$ of positive numbers. This behaviour does not
    contradict~\eqref{eq:met.vx} since~\eqref{eq:met.vx} is only a
    \emph{relative} bound w.r.t.\ a fixed metric $g_v$.
    But~\eqref{eq:vol.vertex} is no longer satisfied.
    
    In addition, the eigenvalue estimate assures that $U_\vxeps$ as
    well as $(U_{v_n})_n$ for a sequence $(v_n)_n \subset V$ do not
    separate into two (or more) parts as $\eps \to 0$ or $n \to
    \infty$, respectively. This could happen e.g.\ if $U_\vxeps$ or
    $(U_{v_n})_n$ consists of two (or more) large parts joined by
    small cylinders which shrink as $\eps \to 0$ resp.\ $n \to
    \infty$.  This could lead to a decoupling of the edges emanating
    from such a vertex or such a sequence of vertices.
  \item We want to stress that $U_v$ cannot either become too small as
    $v \to \infty$ for some sequence of vertices. This is at first
    sight not obvious, since a priori,~\eqref{eq:met.vx} is not a
    restriction of $g_v$ as family in $v \in V$
    and~\eqref{eq:vol.vertex} roughly says that $U_v$ remains small as
    family in $v$. But one has to take into account that $g_\edeps$
    and $g_\vxeps$ are restrictions of a global \emph{smooth} metric
    $g_\eps$. In particular, the metrics on the common boundary $\bd_e
    U_v$ of $U_e$ and $U_v$ must be the same; and therefore, the
    uniform estimates of $g_\edeps$ in~\eqref{eq:asym.met.ed} become
    uniform estimates of $g_\vxeps$ (take e.g.\ a sequence $\rho_n \to
    0$ and argue as in the previous remark).
  \end{enumerate}
\end{remark}
\begin{remark}
  \label{rem:collar.vx}
  Without loss of generality we can assume that $\bd_e U_v$ has a
  collar neighbourhood $U_{e,v}=(0,\ell_-) \times F$ in which the
  metric $g_v$ on $U_v$ has the form
  \begin{equation*}
    g_{e,v}=\dd \overline x^2 + h_{\overline x}
  \end{equation*}
  for $(\overline x,y) \in U_{e,v}$, i.e., the collar neighbourhood
  has length $\ell_-$ for some global constant $0<\ell_-<1$ (e.g.,
  $\ell_-=\ell_0/2$, cf.~also \Fig{graph.emb} where
  $\ell_-<\ell_0/2$).  Here, $\bd_e U_v$ is the part of the boundary
  (diffeomorphic to $F$) where the edge neighbourhood of $e$ meets. In
  addition, we assume that estimates similar to \eqref{eq:met.vol}
  and~\eqref{eq:met.1st.der} are fulfilled (with $\eps=1$, $x$
  replaced by $\overline x$ and $g_\edeps$ replaced by $g_{e,v}$).
  
  If this condition is not satisfied, we just have to modify the
  decomposition of the manifold into edge and vertex neighbourhoods in
  order that $U_\vxeps$ has at least a cylindrical part of length
  $\eps \ell_-$ (taken from the edge neighbourhood). The desired
  estimates \eqref{eq:met.vol} and~\eqref{eq:met.1st.der} follow now
  from~\eqref{eq:asym.met.ed} with the new variable $\overline
  x=x/\eps$.
\end{remark}

\subsection{Main result}
\label{sec:est.graph}
We are now able to prove the following lemmas needed to complete the
proof of the closeness assumptions. Mainly, the estimates are already
given in~\cite{exner-post:05} but since there, we only considered
compact graphs and compact manifolds, a precise control of the
constants was not necessary. We do not repeat the proofs given there
but we state the necessary results together with their dependence on
the constants given in the uniformity assumptions.

\begin{lemma}
  \label{lem:rest0}
  We have
  \begin{equation*}
    \normsqr[\bd_e U_v] u \le 
    c_{\mathrm{tr}} \big( \normsqr[U_v] u + \normsqr[U_v] {du} \big)
  \end{equation*}
  for all $u \in \Sob {U_v}$ where
  \begin{equation*}
    c_{\mathrm{tr}} :=
    \frac{2p_+(1+o_1(1))(1+O_3(1))}{p_-(1-o_1(1)) \ell_-}.
  \end{equation*}
\end{lemma}
\begin{proof}
  The estimate is an immediate consequence of~\eqref{eq:point.vx} and
  \Rem{collar.vx}.
\end{proof}

The following lemma roughly states that the transversal average on the
boundary $\bd_e U_v$ is close to the total average on $U_v$. The proof
is based on the fact that the second Neumann eigenvalue of $U_\vxeps$
tends to $\infty$ as $\eps \to 0$
(cf.~\cite[Lemma~5.5]{exner-post:05}):
\begin{lemma}
  \label{lem:rest}
  The inequality
  \begin{equation*}
    \eps^m |C_v u - N_e u(v)|^2 \le 
    \widetilde c_{\mathrm{tr}} \eps^{2\alpha - 1} \normsqr[U_\vxeps] {d u}
  \end{equation*}
  holds for all functions $u \in \Sob {U_\vxeps}$ and $v =
  \bd_\pm e$ where
  \begin{equation*}
    \widetilde c_{\mathrm{tr}} :=
        \frac{c_-^2}{p_- (1-o_1(1)) c_+^d} 
             \Big( \frac 1 {\lambda_2} + 1 \Big)
              c_{\mathrm{tr}}.
  \end{equation*}
\end{lemma}
Note that $\eps^{2\alpha - 1} \to 0$ as $\eps \to 0$ since $\alpha >
(d-1)/d \ge 1/2$.

The next lemma assures that higher transversal modes does not
contribute too much (cf.~\cite[Lemmas~3.1 and~4.4]{exner-post:05}).
Essentially, it is the observation, that $N_e u - u$ is the projection
onto the orthogonal complement of the first (constant) eigenfunction
on $F$:
\begin{lemma}
  \label{lem:ed.av}
  We have
  \begin{equation*}
    \normsqr[U_\edeps] {N_e u - u} \le 
    c_{\mathrm {ed}} o_4(\eps) \normsqr[U_\edeps] {d u}
  \end{equation*}
  for all $u \in \Sob {U_\edeps}$ where
  \begin{equation*}
    c_{\mathrm {ed}} := \frac{(1+o_1(1))}{(1-o_1(1))\EW 2 F}.
  \end{equation*}
\end{lemma}

Finally, we need to assure that there is no concentration at the
vertex neighbourhoods in any bounded spectral interval
(cf.~\cite[Corollary~5.8]{exner-post:05}):
\begin{lemma}
\label{lem:vx}
  The estimate
  \begin{equation*}
    \normsqr[U_\vxeps] u \le 
    c_{\mathrm{vx}} \eps^{\alpha d - m} \normsqr[1, \widehat U_\vxeps] u
  \end{equation*}
  holds for all $u$ in $\Sob {\widehat U_\vxeps}$ where $\widehat
  U_\vxeps := U_\vxeps \cup \bigcup_{e \in E_v} U_\edeps$ and
  $c_{\mathrm {vx}}$ depends only on $\ell_0$, $p_\pm$, $o_1(1)$,
  $O_3(1)$, $c_\pm$, $c_{\vol}$, $\lambda_2$ and $\widetilde
  c_{\mathrm{tr}}$.
\end{lemma}

We are now able to estimate the RHS of the closeness
assumptions~\eqref{eq:j.scale.gr}--\eqref{eq:j.bdd.gr}
and~\eqref{eq:j.bdd..gr}. For the first one, we have
\begin{equation*}
  \normsqr{Jf - J_1f} \le 
      \frac{4 c_+^{d/2} c_{\vol}}{\ell_0 p_-}
         \eps^{\alpha d - m} \normsqr[1] f
\end{equation*}
for $f \in \HS_1$ using \Lem{point.vx} and~\eqref{eq:met.vx}. Note
that $\eps^{\alpha d - m} \to 0$ as $\eps \to 0$ due
to~\eqref{eq:est.alpha}. Next, we have
\begin{equation*}
  \normsqr{J' u - J_1' u} \le
     p_+ \tilde c_{\mathrm{tr}} \eps^{2\alpha - 1} 
         \sum_{e \in E} \sum_{v \in \bd e} \normsqr[U_\vxeps] {du} \le
     d_0 \, p_+ \tilde c_{\mathrm{tr}} \eps^{2\alpha - 1} \, \wt{\qf h} (u)
\end{equation*}
using \Lem{rest}, \Lem{cut.off} and \Eq{sum.ed.vx}.
The estimation of~\eqref{eq:j.adj.gr}, i.e.,
\begin{equation*}
  |\iprod {Jf} u - \iprod f {J' u} | \le 
  o_1(1) \norm f \norm u 
\end{equation*}
follows from~\eqref{eq:met.vol}. Similarly,
the estimate~\eqref{eq:j.comm.gr} can be proven by
\begin{equation*}
  |\wt{\qf h}(J_1 f, u) - \qf h(f, J_1' u)| \le
  \Bigl( o(1)  + 
     \Bigl[\frac {2d_0 p_+ \widetilde c_{\mathrm{tr}}} {\ell_0} \Bigr]^{1/2} 
      \eps^{\alpha - 1/2}
  \Bigr) \qf h(f)^{1/2} \wt{\qf h} (u)^{1/2}
\end{equation*}
where $o(1)$ depends on the errors in~\eqref{eq:asym.met.ed}. Again, we
have used \Lem{rest}, \Lem{cut.off} and \Eq{sum.ed.vx}.
Estimate~\eqref{eq:j.inv..gr} follows from
\begin{multline*}
  \normsqr{J J' u - u} =
  \sum_{e \in E} \normsqr[U_\edeps] {N_e u - u} +
    \sum_{v \in V} \normsqr[U_\vxeps] u \\ \le
  c_{\mathrm {ed}} o_4(\eps) \sum_{e \in E} \normsqr[U_\edeps] {d u} +
    c_{\mathrm {vx}} \eps^{\alpha d - m} \sum_{v \in V} 
                          \normsqr[1, \widehat U_\vxeps] {u} \\ \le
  3 (c_{\mathrm {ed}} o_4(\eps) + c_{\mathrm {vx}} \eps^{\alpha d - m})
        \normsqr[1] u
\end{multline*}
where we also used~\eqref{eq:sum.vx.ed} in the second estimate.
Finally, Assumption~\eqref{eq:j.bdd} follows from
\begin{equation*}
  \normsqr{J f} \le (1+ o_1(1) \normsqr f \qquad \text{and} \qquad
  \normsqr{J' u} \le \frac 1 {1 - o_1(1)} \normsqr u
\end{equation*}
and~\eqref{eq:met.vol}. We therefore have proven
\begin{theorem}
  \label{thm:graph}
  Suppose that the metric graph $X_0$ and the family of graph-like
  manifolds $X_\eps$ is given as below and satisfy the uniformity
  conditions~\eqref{eq:deg.bd}--\eqref{eq:vol.vertex}. Then the
  generalised weighted Neumann Laplacian on the graph $(\laplacian
  {X_0}, \Lsqr {X_0})$ and the (Neumann) Laplacian on the manifold
  $(\laplacian {X_\eps}, \Lsqr {X_\eps})$ are $\delta$-close of order
  $1$ where $\delta=o(1)$ as $\eps \to 0$.  In particular, all the
  results of \App{abstr.mod} are true, e.g., the convergence of
  eigenfunctions stated in \Thm{eigenvectors} or the spectral
  convergence in \Thm{spectrum}.
\end{theorem}

In particular, we have the following consequence of the convergence
results of \App{abstr.mod}:
\begin{remark}
  Due to \Thm{other.est} and \Thms{cont}{meas}, we can approximate the
  complicated operator on the manifold $\phi(\laplacian{X_\eps})$ by
  the simpler operator $J\phi(\laplacian{X_0})J'$ up to an error,
  where e.g.\ $\phi(\lambda)=(\lambda+1)^{-1}$ (resolvent),
  $\phi_t(\lambda)=\e^{-t\lambda}$ (heat operator) or $\phi=\1_I$
  (spectral projection).  Saito~\cite{saito:00} obtained a similar but
  weaker assertion in the resolvent case.
\end{remark}

\section{Examples and applications of the spectral convergence}
\label{sec:ex.app}
In this section we give several classes of examples for uniform graphs
and manifolds. We also provide consequences of the above-mentioned
spectral convergence.

\subsection{Embedded graphs and graph-like manifolds}
\label{sec:emb.graph}

Let us start with an \emph{embedded} metric graph as an explicite
example in order to illustrate the geometric meaning of the uniformity
assumptions~\eqref{eq:deg.bd}--\eqref{eq:vol.vertex}.  This situation
has originally been treated~in~\cite{kuchment-zeng:01,
  rubinstein-schatzman:01} on compact metric graphs (cf.\ 
also~\cite[Ex.~4.2]{exner-post:05})), and on trees in~\cite{saito:00}:

Suppose that the weighted metric graph $X_0=(V,E,\bd,\ell,p)$ is
isometrically embedded in $\R^2$ via the maps
$\map{\psi_e}{(0,\ell_e)}{\R^2}$. Then
  \begin{equation*}
    \Psi_\edeps(x,y) := \psi_e(x) + \eps r_e(x) y n_e(x), \qquad
    (x,y) \in (0,\ell) \times [-1/2,1/2] \cong e \times F
  \end{equation*}
  defines a weighted tubular neighbourhood of the edge $e$ considered
  as subset of $\R^2$. Here, $n_e$ is a unit vector field normal to
  the tangent vector field $\dot \psi_e$ and $r_e(x)$ defines the
  radius of the neighbourhood. Let $X_\eps$ be the union of the
  closures of the range of all $\Psi_\edeps$, $e \in E$. Denote by
  \begin{equation*}
    \kappa_e := \dot \psi_1 \ddot \psi_2 - \ddot \psi_1 \dot \psi_2
  \end{equation*}
  the signed curvature of the embedded edge $e$. We assume that there
  are global constants $\ell_0, \beta_0, \kappa_0, r_\pm, \dot r_0>0$
  such that
  \begin{align}
    \label{eq:angle} \tag{E1}
    \measuredangle (e,e') &\ge \beta_0, &
    \tan (\beta_0/2)        &> r_+/\ell_0\\
    \label{eq:curve} \tag{E2}
    && |\kappa_e(x)|         &\le \kappa_0, \quad x \in e\\
    \label{eq:len.rad} \tag{E3}
    \ell_e                &\ge \ell_0, &
    r_e (x)               &\ge r_-, \quad d(x,\bd e) \le \min\{1,\ell_e/2\}\\
    \label{eq:rad.upp} \tag{E4}
    |\dot r_e(x)|         &\le \dot r_0, &
    r_e (x)               &\le r_+, \quad x \in e.
  \end{align}
  Here, $\measuredangle(e,e')$ denotes the angle between the tangent
  vectors $\dot \psi_e$ and $\dot \psi_{e'}$ of the embedded edges at
  the vertex $v$ for $e, e' \in E_v$, $e \ne e'$.
  
  We claim that under these assumptions, $X_\eps$ is a \emph{uniform}
  family of graph-like manifolds\footnote{Of course, there are other
    possibilities how to define $X_\eps$ close to the vertices, which
    still lead to a \emph{uniform} graph-like manifold. One can e.g.\ 
    smoothen the singularities at the boundary $\bd U_\vxeps \cap \bd
    X_\eps$ (cf.~\Fig{graph.emb}). In our case, $\bd X_\eps$ has
    singularities but this does not matter in the proof of~\Thm{graph}
    and \Thm{emb.graph}.} embedded in $\R^2$ associated to the
  \emph{uniform} metric graph $X_0$ with weights $p_e := r_e$:
  \begin{theorem}
    \label{thm:emb.graph}
    Under the assumptions~\eqref{eq:angle}--\eqref{eq:rad.upp} \sloppy
    the metric graph $X_0$ and the associated weighted neighbourhood
    $X_\eps$ satisfy~\eqref{eq:deg.bd}--\eqref{eq:vol.vertex} for $\eps$
    small enough. In particular, the generalised weighted Neumann
    Laplacian on the graph $(\laplacian {X_0}, \Lsqr {X_0})$ and the
    Neumann Laplacian $(\laplacian[\Neu] {X_\eps}, \Lsqr {X_\eps})$
    are $\delta$-close of order $1$ where $\delta=O(\eps^{1/2})$ as
    $\eps \to 0$, and therefore, the convergence results of
    \App{abstr.mod} are true.
  \end{theorem}
  \begin{proof}
    The bounded degree assumption~\eqref{eq:deg.bd} follows
    from~\eqref{eq:angle} with $d_0 = \lceil 2\pi/\beta_0 \rceil$.
    \eqref{eq:length} and \eqref{eq:density} are clear. We decompose
    $X_\eps$ into the edge and vertex neighbourhoods $U_\edeps$ and
    $U_\vxeps$, resp., in the way that
    \begin{equation*}
      U_\edeps=
      \Psi_\edeps\bigl( 
             (\eps \ell_0/2, \ell_e - \eps \ell_0/2) \times F \bigr)
    \end{equation*}
    denotes the edge neighbourhood of $e$ shortened by the amount
    $\eps \ell_0$ belonging to the vertex neighbourhoods of $\bd_-e$
    and $\bd_+e$. A straightforward calculation now shows that the
    metric of $U_\edeps \cong (e \times F, g_\eps)$ is represented by
    \begin{equation}
      \label{eq:met.emb}
      G_\eps =
      \begin{pmatrix}
        [(1+ \eps \kappa_e r_e y)^2 + \eps^2 y^2\dot r_e^2]
                           (1-\eps \ell_0/\ell_e)^2 &
        \eps^2 r_e \dot r_e y (1-\eps \ell_0/\ell_e)\\ 
        \eps^2 r_e \dot r_e y (1-\eps \ell_0/\ell_e) &  \eps^2 r_e^2
      \end{pmatrix}.
    \end{equation}
    The uniformity condition~\eqref{eq:asym.met.ed} is fulfilled due
    to the curvature assumption~\eqref{eq:curve} and due
    to~\eqref{eq:rad.upp} and we obtain
    \begin{align*}
      o_1(1) &= 2r_+ \eps (1+\eps \kappa_0 r_+)^{1/2}, &
      1+o_2(1) &
                =1+\eps C(\kappa_0,r_+),\\
      O_3(1) &= C(\kappa_0, r_+, \dot r_0), &
      o_4(\eps) &= 2 r_+^2 \eps^2
    \end{align*}
    provided $0 < \eps < \eps_0 = \eps_0(\kappa_0,r_+,\dot r_0)$,
    where $C(\cdot)$ depend only on the indicated constants, since,
    e.g.,
    \begin{equation}
      \label{eq:met.dens.emb}
      \det G_\eps^{1/2} = \eps r_e (1+\eps
      \ell_0/\ell_e)|1+\eps \kappa_e r_e y|.
  \end{equation}
    
    A similar calculation on $U_\vxeps \cap \Psi_{\eps,e(e \times
      F)}$, $v \in \bd e$, now using $\overline x = x/\eps$ and $y \in
    F$ as coordinates (as far as possible on $U_\vxeps$) shows
    that~\eqref{eq:met.vx} and~\eqref{eq:est.alpha} are satisfied
    with $\alpha=1$ and $c_\pm$ depending only on $\kappa_0$, $r_\pm$
    and $\dot r_0$. As unscaled set $U_v$ we use the ``straightened''
    version of $\frac 1 \eps U_\vxeps$, i.e., we replace the edge
    parts in $\frac 1 \eps U_{v, \eps}$ by their tangentials at $v$;
    the tubular neighbourhood has width $r_e(0)$ (cf.\ 
    \Fig{graph.emb}).
    \begin{figure}[h]
      \begin{center}
        \begin{picture}(0,0)%
           \includegraphics{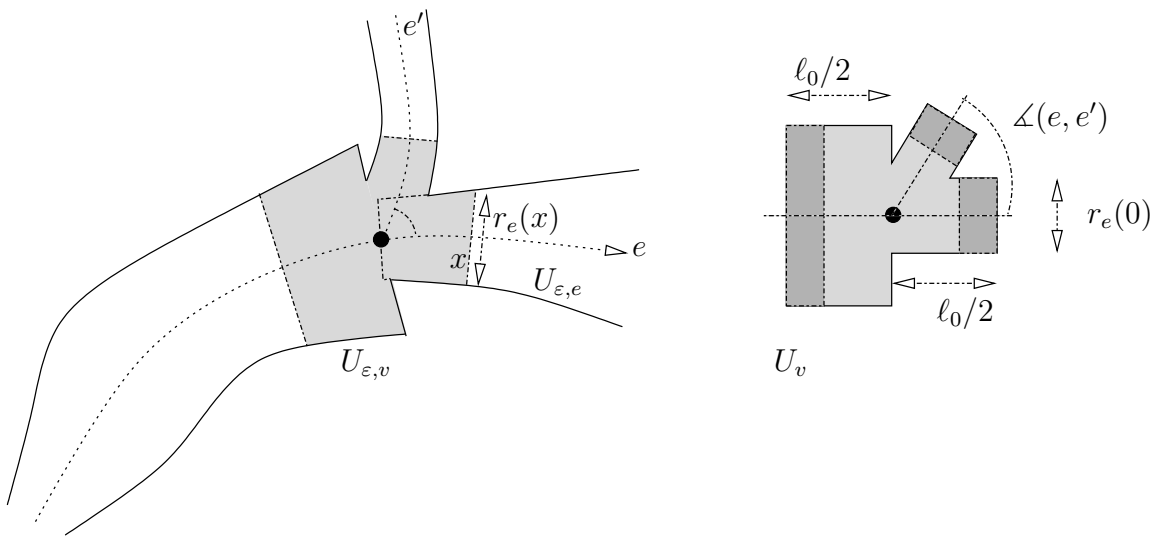}%
        \end{picture}%
        \setlength{\unitlength}{4144sp}%
        \begin{picture}(7103,3171)(98,-2390)
           \put(3849,-709){$e$}%
           \put(4697,-1396){$U_v$}%
           \put(6571,-511){$r_e(0)$}%
           \put(5671,-1096){$\ell_0/2$}%
           \put(4816,344){$\ell_0/2$}%
           \put(6121, 74){$\measuredangle(e,e')$}%
           \put(3015,-536){$r_e(x)$}%
           \put(2757,-776){$x$}%
           \put(2476,614){$e'$}%
           \put(2103,-1396){$U_{\eps,v}$}%
           \put(3256,-909){$U_{\eps,e}$}%
         \end{picture}%
        \caption{Decomposition of the weighted neighbourhood $X_\eps$
          and the unscaled vertex neighbourhood $U_v$. In dark grey we
          denoted the collar neighbourhoods mentioned in
          \Rem{collar.vx}. Note that these rectangles exists due
          to~\eqref{eq:angle}.}
      \label{fig:graph.emb}
      \end{center}
    \end{figure}
    
    Next, the volume estimate of~\eqref{eq:vol.vertex} is satisfied
    due to the angle assumption~\eqref{eq:angle}:
    \begin{equation*}
      \vol U_v \le \sum_{e \in E_v} r_e(v) \ell_0/2 \le
      d_0 r_+ \ell_0/2.
    \end{equation*}
    Due to~\eqref{eq:angle} we can place at each end of $U_v$ a
    rectangle (denoted in dark grey in \Fig{graph.emb}) of length
    $\ell_-=(\ell_0 - r_+ \cot (\beta_0/2))/2>0$ and width $r_e(0) \in
    [r_-,r_+]$. These rectangles are the collar neighbourhoods
    mentioned in \Rem{collar.vx}. The Neumann eigenvalues depend
    continuously on the angles $\measuredangle(e,e') \ge \beta_0$ and
    on the widths $r_e(0) \in [r_-,r_+]$, so~\eqref{eq:vol.vertex}
    follows.
\end{proof}

\begin{remark}
  Although the curvature $\kappa_e$ of the edge and the angle
  $\measuredangle(e,e')$ between two adjacent edges are not detectable
  in the limit (at least not in our first order approximation of an
  eigenvalue $\lambda(\eps)=\lambda(0)+o(1)$), we nevertheless need
  the uniform assumptions~\eqref{eq:angle} and~\eqref{eq:curve}. Using
  the direct eigenvalue estimates of \Rem{ew.better} on compact spaces
  one can show that $\lambda_k(\eps)=\lambda_k(0)+O(\eps^{1/2})$. It
  would be interesting whether one can detect information on the
  curvature or the angles between the edges via an asymptotic
  expansion of $\lambda_1(\eps)$. For a curved tubular neighbourhood
  with Dirichlet boundary conditions around a closed curve of length
  $\ell$ with positive curvature in $\R^3$, the first eigenvalue
  expands as
  \begin{equation*}
    \lambda_1^\Dir(\eps) = 
           \frac{\lambda_1}{\eps^2} - 
           \frac 3{4L} \int_0^\ell \kappa_1(s)^2 \dd s + O(\eps)
  \end{equation*}
  (cf.~\cite[Thm.~4.1]{karp-pinsky:88}) where $\lambda_1$ is the first
  Dirichlet eigenvalue of the unit disc and $\kappa_1$ is the
  curvature of the curve.
\end{remark}
  
\begin{remark}
  \label{rem:horn}
  Note that we do not need a \emph{global} lower bound on the radius
  $r_e$, i.e., infinite edges with a shrinking neighbourhood are
  allowed (e.g.\ horn-like shapes as in~\cite{davies-simon:92}. If the
  spectrum of the Laplacian on the corresponding edge neighbourhood
  $U_\edeps$ is $[0,\infty)$, our analysis does not give new
  information.  More interesting cases are provided if the spectrum on
  $U_\edeps$ is purely discrete, e.g. for radial functions decaying
  fast enough like $r_e(x)=\e^{-x^\beta}$, $\beta>1$
  (cf.~\cite{evans-harris:89, davies-simon:92}). The weighted graph
  Laplacian on $e$ now has the form $(Hf)_e=-f_e''+m \beta x^{\beta-1}
  f_e'$. An example of a horn-like end with infinitely many spectral
  gaps in the essential spectrum was constructed
  in~\cite[Thm.~3]{lott:01}. In principle, these results can be
  recovered with our analysis.
\end{remark}

\subsection{Examples constructed from a finite number of building blocks}
\label{sec:examples}

\subsubsection*{Covering manifolds}
An important class of examples satisfying the uniformity conditions
are coverings with a compact quotient: Clearly, a covering metric
graph $X_0 \to M_0$ with compact quotient $M_0$ is \emph{uniform}.
Similarly, an associated family of graph-like covering manifold
$X_\eps \to M_\eps$ with compact quotient $M_\eps$ is \emph{uniform}
once~\eqref{eq:asym.met.ed}--\eqref{eq:est.alpha} are fulfilled.

For abelian covering groups (and for some classes of non-abelian
\sloppy groups, cf.~\cite{lledo-post:pre04}) the spectral convergence
on \emph{compact} graphs and manifolds would be enough: The Floquet
theory allows to describe the spectrum on the covering via a family of
Laplacians on the compact quotient. Nevertheless our result here is
more general since we can treat an \emph{arbitrary} covering.

More generally, the assumptions
\eqref{eq:asym.met.ed}--\eqref{eq:vol.vertex} are fulfilled if the number
of isomorphism classes of $U_v$ and $U_e$ are finite, i.e., if we
construct the graph-like manifold out of a finite number of building
blocks as in a plumber's shop. An example is given in \Sec{disc.graph}
where we construct a graph-like manifold according to the
Sierpi\'nski graph (cf.~\Fig{sierpinski}).  Here, only two different
vertex neighbourhood building blocks are necessary. Note that this
manifold has a \emph{fractal} structure, not locally, but
\emph{globally}.

\subsubsection*{Spectral gaps and eigenvalues in gaps}

Typically, operators on coverings have a \emph{band-gap} type
spectrum, i.e., the spectrum is the locally finite union of compact
intervals (maybe reduced to a point). The spectral convergence ensures
e.g.\ the existence of spectral gaps as $\eps \to 0$ once $\spec
{\laplacian {X_0}}$ has spectral gaps and $\eps$ is small
enough:
\begin{theorem}
  \label{thm:gaps} Suppose that $M_0$ is a compact graph with
  associated uniform graph-like manifold $M_\eps$. Denote by $X_0$
  resp.\ $X_\eps$ a covering of $M_0$ resp.\ $X_\eps$ such that
  $X_\eps$ is a graph-like manifold associated to $X_0$. If the
  generalised Neumann Laplacian on $X_0$ has a spectral gap $(a,b)$,
  i.e., $\spec {\laplacian{X_0}} \cap (a,b) = \emptyset$, then the
  Laplacian $\laplacian{X_\eps}$ has a spectral gap close to $(a,b)$
  provided $\eps$ is small enough.
\end{theorem}
Covering metric graphs $X_0 \to M_0$ with spectral gaps are given
e.g.~in~\cite[Sec.~9.4--9.6]{exner-post:05}; the simplest example is
maybe the Cayley graph of the group $\Gamma = \Z \times \Z_p$ where
$\Z_p$ denotes the cyclic group of order $p$ has a spectral gap iff
$p$ is odd. A similar example consists of a regular rooted tree
(cf.~\Thm{tree.mfd}). A different procedure of creating gaps in a
metric graph is provided by the so-called \emph{graph-decoration}.
Roughly speaking, the new graph $\hat X_0$ is obtained from a given
infinite graph $X_0$ by attaching a fixed (compact) graph $M_0$ to
each vertex $v$ of $X_0$. The Laplacian on $\hat X_0$ now has spectral
gaps. This result has been established for a discrete graph
in~\cite{aizenman-schenker:00}. The general case for quantum graphs is
announced in~\cite{kuchment:05} and proved in the case when $X_0$ is a
compact graph (in the sense that there is no spectrum of the decorated
graph near (certain parts) of the spectrum of the decorating graph).
For quantum graphs, there are related examples leading to gaps
(cf.~\cite{ael:94}, \cite{exner:95}). A similar effect by attaching a
single loop to each vertex of a periodic graph has been used
in~\cite[Sec.~9]{exner-post:05}. The case of periodically arranged
manifolds connected by line sements or through points has been
analysed in~\cite{beg:03, bgl:05}.

Finally, another class of examples is given by \emph{fractal} metric
graphs, i.e., graphs arranged in a self-similar manner
(cf.~\Thm{sierpinski}). The main point here is that the metric graph
spectrum has a fractal structure, so once, $\eps$ is small enough, the
corresponding Laplacian on a graph-like manifold has an arbitrary (but
a priori finite) number of spectral gaps \emph{in the compact}
spectral interval $[0,\Lambda]$.

\subsubsection*{Eigenvalues in gaps}
Suppose that $X_0$ is a uniform metric graph such that its generalised
Neumann Laplacian has a spectral gap, namely, $\spec {\laplacian
  {X_0}} \cap (a,b) = \emptyset$. By the previous example, a
corresponding \emph{uniform} graph-like manifold $X_\eps$ has also a
spectral gap close to $(a,b)$ provided $\eps$ is small enough.
  
Now if we change the metric graph locally, e.g.\ by attaching a loop
of length $\ell_1$ at a fixed vertex $v_1 \in V$ (call the perturbed
graph $\hat X_0$) then the generalised Neumann Laplacian on $\hat X_0$
has additional eigenvalues $\lambda_k=(2\pi k/\ell_1)^2$ with
eigenfunctions located on the loop and vanishing at $v_1$ and on the
rest of the graph (maybe there are more additional eigenvalues). For
example, if $2\pi/\sqrt b < \ell_1 < 2\pi/\sqrt a$ then the ground
state of the loop lies in $(a,b)$.  Note that the essential spectrum
of $X_0$ and the perturbed metric graph $\hat X_0$ are the same since
the perturbation is compact; in particular, only \emph{discrete}
eigenvalues occur in the spectral gap. Due to the spectral convergence
\Thm{graph}, a corresponding (uniform) graph-like manifold now must
have (at least) an eigenvalue in $(a,b)$.  Furthermore, if the
corresponding eigenvalue of the quantum graph is simple (i.e., there
are no other eigenvalues of the looped graph $\hat X_0$ at
$\lambda_1=(2\pi/\ell_0)^2$), then there is a unique eigenvalue close
to $\lambda_1$ of multiplicity $1$, and the corresponding
eigenfunction of the graph-like manifold is close to the eigenfunction
of the metric graph in the sense of \Thm{eigenvectors}. More
generally, we have:
\begin{theorem}
  \label{thm:ew.graph}
  Suppose that $\laplacian {X_0}$ has a discrete eigenvalue $\lambda$
  of multiplicity $m$ and $\spec {\laplacian {X_0}} \cap I =
  \{\lambda\}$. Let $X_\eps$ be a graph-like manifold associated to
  $X_0$. Then $\laplacian {X_\eps}$ has $m$ eigenvalues (not all
  necessarily distinct) converging to $\lambda$ as $\eps \to 0$ and
  the eigenprojections converge in norm, i.e., $\norm{J'P_\eps J -
    P_0} \to 0$ as $\eps \to 0$ where $P_\eps$ is the eigenprojection
  of $\laplacian {X_\eps}$ onto the interval $I$, $\eps \ge 0$. 
  
  If in particular, $\lambda$ is a simple eigenvalue with
  eigenfunction $\phi$, then there exists an eigenfunction $\phi_\eps$
  of $\laplacian {X_\eps}$ such that $\norm {J\phi - \phi_\eps} \to
  0$.
\end{theorem}
Roughly speaking, the theorem says, that an eigenfunction on the
graph-like manifold is approximately given by the corresponding
eigenfunction on the graph, extended constantly in the transversal
direction (and set to $0$ in the vertex neighbourhood).

Eigenvalues in gaps have been discussed e.g.~in~\cite{aadh:94,
  post:03b} (see also the references therein). One can interprete such
a local perturbation as an \emph{impurity} of a periodic structure,
say, a crystal or a semi-conductor. The additional eigenvalue in the
gap now corresponds to an additional energy level and a bound state.
Note that in~\cite{post:03b} it was \emph{assumed} that $X_\eps$ is a
\emph{covering} manifold and that it fulfilled a \emph{gap} condition
in the sense that there is a fundamental domain $D$ such that $\EWD k
D < \EWN {k+1} D$ for some $k$ (Dirichlet and Neumann eigenvalues on
$D$). This is in general a stronger condition than just the assumption
that $\laplacian {X_\eps}$ has a spectral gap.  Here, we only need to
know that the generalised Neumann Laplacian $\laplacian {X_0}$ on the
graph has a spectral gap which is often easy to show.

\subsection{Equilateral metric graphs and discrete graphs}
\label{sec:disc.graph}
In this subsection we will analyse the special case when all length of
the metric graph $X_0=(V,E,\bd, \ell)$ are the same, say
$\ell_e=\ell$. Under these assumptions, there is a nice relation
between the spectrum of the metric graph Laplacian $\laplacian {X_0}$
(with weights $p_e=1$) acting on $\Lsqr{X_0} = \oplus_e \Lsqr{0,\ell}$
and the discrete Laplacian $\laplacian G$ of the graph $G=(V,E,\bd)$.
Roughly speaking, the spectrum of $\laplacian {X_0}$ consists of
infinitely many (distorted) copies of the spectrum of $\laplacian G$
arranged in a row.  This relation allows us to profit from the vast
literature on $\laplacian G$ and to carry over calculations of spectra
of the discrete to the continuous graph Laplacian.

The \emph{discrete} Laplacian is given by
\begin{equation}
  \label{eq:def.disc.lap}
  (\laplacian G a)(v) := -\frac 1 {\deg v} \sum_{w \sim v}
  (a(w)-a(v)),
  \qquad a \in \lsqr V
\end{equation}
where $w \sim v$ means that $v$ and $w$ are joined by an edge. The
discrete Laplacian acts in the weighted space $\lsqr V$ consisting of
all sequences $a=(a_v)_v$ with finite weighted norm
\begin{equation*}
  \norm[\lsqr V] a := \sum_{v \in V}{\deg v}|a(v)|^2.
\end{equation*}
This operator is bounded and has spectrum in $[0,2]$.

In~\cite{cattaneo:97} one can find the following nice relation between
the spectrum of the generalised Neumann Laplacian $\laplacian {X_0}$
and the discrete Laplacian $\laplacian G$. The Dirichlet spectrum
$\Sigma^\Dir = \set{ (k\pi/\ell)^2}{k \in \N}$ of an individual edge
$e \cong (0,\ell)$ always plays a special role. Since we are only
interested in qualitative results (e.g.\ the fractal example in
\Thm{sierpinski}) we exclude the Dirichlet spectrum here. A more
detailed discussion on the Dirichlet spectrum can be found
in~\cite{kuchment:05} and~\cite{cattaneo:97}. We set
$g(\lambda)=1-\cos(\ell \sqrt \lambda)$. Note that $\Sigma^\Dir =
g^{-1}\{0,2\}$.

\begin{theorem}
  \label{thm:disc.met}
  Assume that $G=(V,E,\bd)$ is a countable, connected graph with $\deg
  v \in \{2, 3, \dots, d_0\}$, $v \in V$, and without self-loops then
  \begin{equation*}
    \spec[\bullet]{\laplacian {X_0}} \setminus \Sigma^\Dir =
    g^{-1}\bigl(\spec[\bullet]{\laplacian G} \setminus \{0, 2\}\bigr),
  \end{equation*}
  i.e., for $\lambda \ne \Sigma^\Dir$, we have $\lambda \in
  \spec[\bullet]{\laplacian {X_0}}$ iff $g(\lambda) \in
  \spec[\bullet]{\laplacian G}$. Here $\bullet \in \{\mathrm p,
  \mathrm c, \emptyset, \mathrm {pp}, \mathrm{disc}, \mathrm{ess}\}$
  denotes either the point spectrum (the set of eigenvalues
  $\sigma_{\mathrm p}$), the complement of the
  eigenvalues\footnote{Note that $\sigma_{\mathrm c}$ does in general
    not coincide with the continuous spectrum,
    cf.~\cite{reed-simon-1}.} ($\sigma_{\mathrm c}= \sigma \setminus
  \sigma_{\mathrm p}$), the entire spectrum ($\sigma$), the pure point
  spectrum (the closure of the set of eigenvalues, i.e.,
  $\sigma_{\mathrm{pp}}=\clo{\sigma_{\mathrm p}}$), the discrete or
  the essential spectrum.  Furthermore, the eigenvalue $\lambda$ of
  $\laplacian {X_0}$ has multiplicity $m$ iff the eigenvalue
  $g(\lambda)$ of $\laplacian G$ has multiplicity $m$.
\end{theorem}
\begin{proof}
  The assertion has been proved for the point spectrum and its
  complement in~\cite{cattaneo:97} and therefore also for the entire
  spectrum. Since $g$ is a local homeomorphism on the complement of
  $\Sigma^\Dir$, the statement on $\sigma_{\mathrm{pp}}$ also
  follows. In addition, note that
  \begin{gather*}
    \map {U_\lambda} {E_{g(\lambda)}(\laplacian G)} 
                     {E_\lambda(\laplacian{X_0})},\\
    (U_\lambda a)_e(x) := 
        \frac {\sqrt \ell} {\sqrt 2 \sin (\ell \sqrt \lambda)} \bigl( a(\bd_-e)
    \sin ((\ell - x) \sqrt \lambda) + a(\bd_+e) \sin (x \sqrt \lambda) \bigr)
  \end{gather*}
  is an isometry from the eigenspace of $\laplacian G$ w.r.t.\ the
  eigenvalue $g(\lambda)$ onto the eigenspace of $\laplacian {X_0}$
  w.r.t.\ the eigenvalue $\lambda \notin \Sigma^\Dir$.  In particular,
  multiplicities of eigenvalues are preserved. Furthermore, $\lambda$
  is isolated in $\spec{\laplacian {X_0}}$ iff $g(\lambda)$ is
  isolated in $\spec{\laplacian G}$, so the statement on the discrete
  spectrum follows. For the essential spectrum note that we already
  have the statement for the entire spectrum and for the discrete
  spectrum, and that $\sigma_{\mathrm{ess}}=\sigma \setminus
  \sigma_{\mathrm{disc}}$.
\end{proof}

Since the spectrum of the discrete Laplacian has been explored in many
cases, the previous theorem allows us to determine the corresponding
spectrum of the equilateral metric graph leading to interesting
examples of graph-like manifolds. For simplicity, we fix the edge
length to $\ell=1$.

\subsubsection*{Homogeneous trees}
Let $G$ be a homogeneous routed tree of degree $d_0 \ge 3$, then
  \begin{equation}
    \label{eq:sp.tree}
    \spec[\mathrm p]{\laplacian G} = \emptyset \qquad\text{and}\qquad
    \spec[\mathrm c]{\laplacian G} =
      \left[ 1 - \frac{2\sqrt{d_0-1}} {d_0}, 
             1+\frac{2\sqrt{d_0-1}} {d_0} \right].
  \end{equation}
  \Thm{disc.met} now describes $\spec[\mathrm c]{\laplacian{X_0}}$.
  Since $\spec{\laplacian G} = \spec[\mathrm c]{\laplacian G}
  \subsetneq [0,2]$, the metric graph Laplacian has spectral gaps
  (i.e., no spectrum at) $I_0:=(0, \omega_0^2)$ and
\begin{equation*}
    I_k:=\bigl((k\pi - \omega_0)^2, (k\pi)^2 \bigr) \cup
         \bigl((k\pi)^2, (k\pi + \omega_0)^2\bigr), \qquad k \in \N,
\end{equation*}
where $\omega_0=\arccos (2\sqrt{d_0-1}/d_0)$. A more detailed analysis
done in~\cite{cattaneo:97} shows that $\spec[\mathrm p]{\laplacian
  {X_0}}=\Sigma^\Dir$. All eigenvalues have infinite multiplicity, so
$\spec {\laplacian {X_0}}$ is purely essential and has band-gap
structure with gaps exactly at $I_k$, $k \in \N_0$. In particular,
$\inf \spec {\laplacian {X_0}} = \omega_0^2>0$ and we have:
\begin{theorem}
  \label{thm:tree.mfd}
  Suppose $X_\eps$ is a family of uniform graph-like manifolds
  associated to the regular tree of degree $d_0 \ge 3$ with equal edge
  lengths. Then the Laplacian on $X_\eps$ has spectral gaps and its
  essential spectrum is non-empty near any $\lambda \in \essspec
  {\laplacian {X_0}}$ provided $\eps$ is small enough
  (cf.~\Cor{ess.sp}). Furthermore, $\Sigma_0(X_\eps):=\inf \spec
  {\laplacian{X_\eps}} \to \omega_0^2$ as $\eps \to 0$ and in
  particular, $\Sigma_0(X_\eps) \ge c>0$ for $\eps$ small enough.
\end{theorem}
\begin{remark}
  Note that if $X_\eps$ would be a Riemannian covering of a compact
  manifold $M_\eps \cong X_\eps /\Gamma$, then the non-amenability of
  $\Gamma$ (if e.g.~$\Gamma$ contains the non-abelian free group
  $\Z^{*2}$ as subgroup) implies $\Sigma_0(X_\eps)>0$
  (cf.~\cite{brooks:81}). An \emph{unrooted} tree of degree $d_0$ can
  be considered as \emph{unoriented} Cayley graph of the group
  $\Z_2^{*d_0}$ (free product of $d_0$ groups of order $2$) with
  respect to the generators $\gamma_i$, $i=1, \dots, d_0$. But the
  elements of order $2$ act on a corresponding graph-like manifold as
  reflections. In particular, these elements have fixed points and the
  quotient is no longer a smooth manifold so we cannot use the
  covering argument here in order to show $\Sigma_0(X_\eps)>0$.
\end{remark}

\begin{remark}
  \label{rem:brooks}  
  There is a nice upper estimate for $\Sigma_{\mathrm{ess}}(X):=\inf
  \essspec {\laplacian X}$: Denote by
  \begin{equation*}
    \mu(X) := \lim_{r \to \infty} \frac 1 r \log \vol_d B(x_0,r)
  \end{equation*}
  the growth rate of a ball of radius $r$ in $X$. It can be easily
  seen that $\mu(X)$ is independent of $x_0$. Brooks~\cite{brooks:81b}
  showed that if $\vol_d(X)=\infty$ then
  \begin{equation*}
    \Sigma_{\mathrm{ess}}(X) \le \frac 14 \mu(X)^2.
  \end{equation*}
  A priori, this estimate is not sharp as there are amenable groups
  $\pi$ of exponential growth (cf.~\cite{milnor:68}): For the
  universal cover $\wt M$ of a compact manifold $M$ with amenable
  fundamental group $\pi_1(M)=\pi$ we have $\Sigma_{\mathrm{ess}}(\wt
  M)=0$ but $\mu(\wt M)>0$.
  
  In our case, we can easily calculate $\mu(X_\eps)$ approximately: If
  $x_0$ is a point in the root vertex neighbourhood then an increase
  of $r$ by $\ell_0=1$ encounters another generation in the tree.
  Therefore $\vol_d B(x_0,r) \approx \vol_d \hat U_\vxeps
  ((d_0-1)^r-1)$ for large $r$ where $\hat U_\vxeps$ is a (sample)
  vertex neighbourhood together with the $d_0$ adjacent half edge
  neighbourhoods. We obtain $\mu(X_\eps) \approx \log(d_0-1)$ and in
  particular, our estimate shows
  \begin{equation*}
    \Sigma_0(X_\eps)=
    \Sigma_{\mathrm{ess}}(X_\eps) \approx 
      \arccos^2 \Bigl( \frac{2\sqrt{d_0-1}} {d_0} \Bigr) < 
      \frac 1 4 \log^2(d_0-1) \approx \frac 14 \mu(X_\eps)^2
  \end{equation*}
  if $d_0 \ge 3$. The inequality provides an (approximative)
  \emph{lower} bound on $\mu(X_\eps)$ and the difference of the LHS
  and RHS is e.g.\ smaller than 1\% if $d_0=3$). For further estimates
  of this type using isoperimetric constants we refer
  to~\cite{brooks:81b}. We only want to mention here that
  isoperimetric constants on graph-like manifolds can quite easily be
  calculated.
\end{remark}

\subsubsection*{Sierpi\'nski graphs}
\begin{figure}[h]
  \begin{center}
    \begin{picture}(0,0)%
      \includegraphics{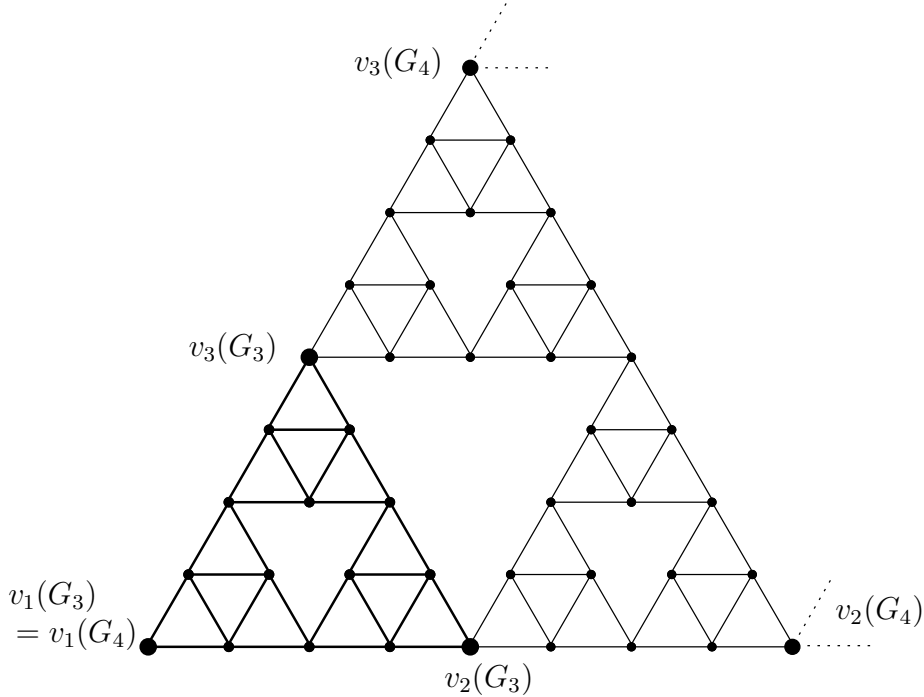}
    \end{picture}%
    \setlength{\unitlength}{4144sp}%
    \begin{picture}(5857,4184)(496,-3481)
      \put(496,-2941){$v_1(G_3)$}%
      \put(1576,-1456){$v_3(G_3)$}%
      \put(5446,-3031){$v_2(G_4)$}%
      \put(3106,-3436){$v_2(G_3)$}%
      \put(541,-3166){$=v_1(G_4)$}%
      \put(2566,254){$v_3(G_4)$}%
    \end{picture}%
  \caption{The first four generations $G_4$ of the infinite Sierpi\'nski
    graph, each edge having unit length. The graph $G_3$ is denoted
    with thick edges and is naturally embedded into $G_4$.}
  \label{fig:sierpinski}
\end{center}
\end{figure}

Let us give another interesting example admitting a fractal spectrum.
Let $G_1$ be the complete graph with three vertices. Suppose
$G_n=(V_n,E_n)$ has already been constructed. Then $G_{n+1}$ is
obtained from three disjoint copies $G_n^{(i)}$ ($i=1,2,3$) of $G_n$
and the equivalent relation $\sim$, i.e., $G_{n+1} := G_n^{(1)} \dcup
G_n^{(2)} \dcup G_n^{(3)}/{\sim}$ where $\sim$ identifies the vertex
$v_i(G_n^{(j)})$ with $v_j(G_n^{(i)})$, $i\ne j$, $i,j=1,2,3$. Here,
$v_1(G_n)$ denotes the lower left vertex of degree $2$, $v_2(G_n)$ the
lower right vertex of degree $2$ and $v_3(G_n)$ the upper vertex of
degree $2$ (cf.~\Fig{sierpinski}). Note that each $G_n$ has exactly
three vertices of degree $2$, the other have degree $4$. Furthermore,
$|V_n|=(3^n+1)/2$ and $|E_n|=3^n$. Now, $G_n$ embeds into $G_{n+1}$
via $G_n = G_n^{(1)} \hookrightarrow G_{n+1}$. Finally,
the Sierpi\'nksi graph is given by $G := \bigcup_{n \in \N} G_n$
(cf.~\Fig{sierpinski}).
Note that $G$ has one vertex of degree $2$, all other vertices have
degree $4$. The nature of the spectrum of the discrete Laplacian
$\laplacian G$ was calculated by~\cite[Thm.~2]{teplyaev:98}: We have
\begin{equation}
  \label{eq:spec.sier}
  \spec {\laplacian G} = J \cup D, \qquad
  D := \Bigl\{\frac 32\Bigr\} \cup 
       \bigcup_{n=0}^\infty p^{-n} \Bigl\{\frac 34 \Bigr\},
  \qquad
  J := \clo D,
\end{equation}
where $p(z):=z(5-4z)$. and $p^{-n}$ is the $n$-th pre-image, i.e.,
$p^{-n}\{3/4\} = \set{z \in \R}{p^{\circ n}(z)=3/4}$. The spectrum of
$\laplacian G$ is pure point, each eigenvalue has multiplicity
$\infty$, so the spectrum is also purely essential. The set $D$
consists of the isolated eigenvalues and the set $J$ is a Cantor set
of Lebesgue measure $0$ (the Julia set of the polynomial $p$). Due to
\Thm{disc.met} the spectrum of the metric graph Laplacian
$\laplacian{X_0}$ (with $\ell=1$) is pure point and purely essential.
It is given by
\begin{equation*}
  \spec {\laplacian {X_0}} \setminus \Sigma^\Dir =
  g^{-1}(D) \cup g^{-1}(J \setminus \{0\}).
\end{equation*}
Since $g(\lambda)=1-\cos \sqrt \lambda$ is a local homeomorphism
mapping measure $0$ set into measure $0$ sets and vice versa,
$g^{-1}(D)$ consists of isolated eigenvalues of $\laplacian {X_0}$ of
infinite multiplicity and $g^{-1}(J \setminus \{ 0\})$ is a Cantor set
of measure $0$ (away from $\Sigma^\Dir$).

Now our spectral convergence result on graph-like manifolds leads to
an example of a family of smooth manifolds approaching a fractal
spectrum:
\begin{theorem}
  \label{thm:sierpinski}
  Suppose $X_\eps$ is a family of uniform graph-like manifolds
  associated to the Sierpi\'nski graph with equal edge lengths. Then
  the essential spectrum of the Laplacian on $X_\eps$ approaches the
  fractal spectrum of $\laplacian {X_0}$ in any fixed interval
  $[0,\Lambda]$. The discrete spectrum of $\laplacian{X_\eps}$ is
  either empty or merges into the essential spectrum as $\eps \to 0$
  (cf.~\Cor{ess.sp}). In particular, $\spec {\laplacian {X_\eps}}$ has
  an arbitrary large number of spectral gaps in the \emph{compact}
  interval $[0,\Lambda]$ provided $\eps$ is small enough.
\end{theorem}

\subsection{The decoupling case}
\label{sec:decoupling}

We obtain similar convergence results on \emph{non-compact} graph-like
manifolds $X_\eps$ with slower decay of $\vol U_\vxeps$ (i.e., large
vertex neighbourhoods $U_\vxeps$) as discussed in the compact case
in~\cite{kuchment-zeng:03} and~\cite[Sec.5--8]{exner-post:05}.  If for
example the vertex volume $\vol U_\vxeps$ decays slower than the edge
volume $\vol U_\edeps$ then the Laplacian on $X_\eps$ converges also
in the non-compact case to the \emph{decoupled} limit operator $H :=
\bigoplus_e H_e \oplus 0$ acting in the enlarged Hilbert space $\HS :=
\Lsqr X_0 \oplus \C^V$.  Here, $H_e$ is the Dirichlet operator on the
interval $(0,\ell_e) \cong e$ and $0$ is the null operator on $\C^V$.
Therefore, the limit operator has pure point spectrum
\begin{equation}
  \label{eq:spec.dec}
    \spec H = \Bigset{\lambda_k(e):=\frac {\pi^2 k^2} {\ell_e^2}} 
                  {e \in E, k \in \N_0}
\end{equation}
and the multiplicity of $\lambda$ is $|V|$ if $\lambda=0$ and $|\set{e
  \in E}{\lambda=\lambda_k(e)}|$ if $\lambda>0$. We omit the details
here since the proof is rather similar to the non-decoupling case
treated below. Note that we need here some uniformity assumptions on
the vertex neighbourhood $U_\vxeps$ ensuring that $U_\vxeps$ does not
become too \emph{small}, e.g., we need a \emph{lower} bound on $\vol
U^-_\vxeps$.  Here $U^-_\vxeps$ is the subset of $U_\vxeps$ where the
metric satisfies $g_\vxeps \cong \eps^{2\alpha} g_v$ and $0 < \alpha <
(d-1)/d$ (cf.~\Rem{unif.mfd}.\ref{rem:bottle.neck}
and~\ref{rem:unif.mfd}.\ref{rem:critical}).

\begin{theorem}
  \label{thm:decoupling}
  Suppose that $X_\eps$ is a graph-like manifold with large vertex
  neighbourhoods associated to a metric graph $X_0$ with edge length
  $(\ell_e)_e$. Then the Laplacian on $X_\eps$ (with Neumann boundary
  conditions, if $\bd X_\eps \ne \emptyset$) approaches the decoupled
  operator $H$. In particular, its spectrum converges to the set
  $\spec H$ given in~\eqref{eq:spec.dec} in any compact spectral
  interval.
\end{theorem}
  
In a similar way, the case of Dirichlet boundary conditions and small
junctions as in~\cite{post:05a} can be extended to the non-compact
case. The Laplacian on $X_\eps$ with Dirichlet boundary conditions has
to be shifted by $\lambda_1^\Dir(F)/\eps^2$ where $\lambda_1^\Dir(F)$
is the first Dirichlet eigenvalue of the cross section. The limit
operator also decouples and consists of the Dirichlet eigenvalues
only.

Since the graph structure is no more visible in the limit due to the
decoupling, we just give a simple example of a graph consisting of a
half-line with one vertex of degree $1$ at $0$ and the others of
degree $2$.  We label the edges by $n \in \N$. By an appropriate
choice of the length $\ell_e$ we can construct a manifold with certain
spectral properties. If we consider e.g.\ $\ell_n:=\sqrt n/\pi$ then
$\lambda_k(n)=k^2/n$. Since every rational number $r$ can be written
in the form $r=k^2/n$ ($r=p/q=p^2/(pq)$), the operator $H$ has dense
point spectrum consisting of all non-negative rational numbers.
Therefore the spectrum of $H$ is $[0,\infty)$ and purely essential.
Our analysis here is too weak to say anything more on the nature of
the Laplacian on the graph-like manifold $X_\eps$ with large vertex
neighbourhoods than it is purely essential in any bounded spectral
interval. We expect that $\laplacian {X_\eps}$ also has pure point
spectrum, but one needs arguments from scattering theory to prove
this.

Other choices of the length are possible, e.g., a set of rationally
independent length $\ell_e$, $e \in E$. Once one is able to determine
$\spec H$ by number theoretical arguments one has a Laplacian
$\laplacian {X_\eps}$ with a spectrum close to this set.

\setcounter{section}{0}
\renewcommand{\thesection}{\Alph{section}}
\section{Appendix}
\label{sec:abstr.mod}
In the appendix we prove our main technical tool, the convergence
results for arbitrary pairs of self-adjoint non-negative operators and
Hilbert spaces $(H,\HS)$ and $(\wt H, \wt \HS)$ being close to each
other (cf.~\Def{closeness}). Although most of the techniques are
standard, we repeat the arguments here since usually, one has a
\emph{fixed} Hilbert space and a careful analysis of the dependence on
some parameter entering in the operator \emph{and} Hilbert space is
not necessary. We do not mention the parameter explicitely but express
all convergence results in terms of a ``distance'' $\delta$ of
$(H,\HS)$ and $(\wt H,\wt \HS)$.

\subsection{Scale of Hilbert spaces associated with a non-negative
  operator}
\label{sec:scale}

To a Hilbert space $\HS$ with inner product $\iprod \cdot \cdot$ and
norm $\norm \cdot$ together with a non-negative, unbounded, operator
$H$, we associate the scale of Hilbert spaces
\begin{equation}
  \label{eq:scale.hs}
  \HS_k := \dom (H+1)^{k/2}, \qquad 
    \norm[k] u := \norm{(H+1)^{k/2}u}, \quad k \ge 0.
\end{equation}
For negative exponents, we define
\begin{equation}
  \label{eq:scale.hs.neg}
  \HS_{-k} := \HS_k^*.
\end{equation}
Note that $\HS = \HS_0$ embeds naturally into $\HS_{-k}$ via $u \mapsto
  \iprod u \cdot$ since
\begin{displaymath}
  \norm[-k] {\iprod u \cdot} =
  \sup_{v \in \HS_k} \frac{|\iprod u v|}{\norm[k] v} =
  \sup_{w \in \HS_0} \frac{|\iprod {R^{k/2}u} w|}{\norm[0] w} =
  \norm[0]{R^{k/2} u},
\end{displaymath}
where
\begin{equation}
  \label{def:res}
  R := (H+1)^{-1}
\end{equation}
denotes the resolvent of $H \ge 0$. The last equality used the natural
identification $\HS \cong \HS^*$ via $u \mapsto \iprod u \cdot$.
Therefore, we can interprete $\HS_{-k}$ as the completion of $\HS$ in
the norm $\norm[-k] \cdot$. With this identification, we
have
\begin{equation}
  \label{eq:norm.iprod}
  \norm[-k] u =  \sup_{v \in \HS_k} \frac{|\iprod u v|}{\norm[k] v}, \qquad
  \text{for all $k \in \R$.}
\end{equation}
For a second Hilbert space $\wt \HS$ with inner product $\iprod \cdot
\cdot$ and norm $\norm \cdot$ together with a non-negative, unbounded,
operator $\wt H$, we define in the same way a scale of Hilbert spaces
$\wt \HS_k$ with norms $\norm[k] \cdot$.

Guided by the classical application $H=\laplacian X$ in $\HS=\Lsqr X$
for a complete manifold $X$, we call $k$ the \emph{regularity order}.
In this case, $\HS_k$ corresponds to the $k$-th Sobolev space
$\Sob[k]X$.

\subsection{Operators on scales}
\label{sec:op}

Suppose we have two scales of Hilbert spaces $\HS_k$, $\wt \HS_k$
associated to the non-negative operators $H$, $\wt H$ with resolvents
$R := (H+1)^{-1}$, $\wt R := (\wt H+1)^{-1}$, respectively. The norm
of an operator $\map A {\HS_k} {\wt \HS_{-\wt k}}$ is
\begin{equation}
  \label{eq:norm.op}
  \norm[k \to -\wt k] A :=
  \sup_{u \in \HS_k} \frac{\norm[-\wt k]{Au}}{\norm[k] u} =
  \norm[0 \to 0] {\wt R^{\wt k/2} A R^{k/2}}.
\end{equation}
The norm of the adjoint $\map {A^*} {\wt \HS_{\wt k}} {\HS_{-k}}$ is given by
\begin{equation}
  \label{eq:norm.adj}
  \norm[\wt k \to -k] {A^*} =
  \norm[k \to -\wt k] A.
\end{equation}
Furthermore,
\begin{equation}
  \label{eq:norm.op.ineq}
  \norm[k \to -\wt k] A \le
  \norm[m \to -\wt m] A \qquad \text{provided} \qquad k \ge m,  \wt k \ge \wt m
\end{equation}
since
\begin{displaymath}
  \norm[k \to -\wt k] A =
  \norm[0 \to 0] {\wt R^{\wt k/2} A R^{k/2}} =
  \norm[0 \to 0] {\wt R^{(\wt k - \wt m)/2} \wt R^{\wt m/2} A 
                    R^{m/2} R^{(k-m)/2}} \le
  \norm[m \to -\wt m] A
\end{displaymath}
and $\norm R, \norm {\wt R} \le 1$.

\subsection{Closeness assumption}
\label{sec:ass}

In this section we state our main assumptions on the two operators $H$
and $\wt H$ acting in the Hilbert spaces $\HS$ and $\wt \HS$.  We
think of $(\wt H,\wt \HS)$ being a perturbation of $(H,\HS)$, or that
$(H,\HS)$ describes a simplified model (say, on a metric graph $X_0$)
which is close to a more complicated model given by $(\wt H, \wt \HS)$
(say, on a graph-like manifold $X_\eps$).  We want to state
assumptions under which $(H,\HS)$ and $(\wt H,\wt \HS)$ are close in a
sense to be specified in \Def{closeness}.

Let us explain the following concept of \emph{quasi-unitary} operators
in the case of unitary operators (cf.\ also \Ex{nrconv}): Suppose we
have a unitary operator $\map J \HS {\wt \HS}$ with inverse $\map {J'
  = J^*} {\wt \HS} \HS$ respecting the quadratic form domains, i.e.
$\map {J_1:=J \restr {\HS_1}} {\HS_1} {\wt \HS_1}$ and $\map {J_1'=J^*
  \restr {\wt \HS_1}} {\wt \HS_1}{\HS_1}$. If
\begin{displaymath}
  J_1'^* H = \wt H J_1
\end{displaymath}
then $H$ and $\wt H$ are unitarily equivalent and have therefore the
same spectral properties. The main point here is that $J$ respects the
quadratic form domain and therefore, $\map {{J'_1}^*} {\HS_{-1}}
{\wt \HS_{-1}}$ is an extension of $\map J \HS {\wt \HS}$. In this sense, the
above equality says that $J$ is an intertwining operator.

We want to lessen the assumption such that the spectral properties are
not the same but still at close quarters.  We now start with the
definition of $\delta$-closeness:

\begin{definition}
  \label{def:closeness}
  Suppose we have linear operators
  \begin{equation}
  \label{eq:trans.op}
  \begin{array}{ccc}
    \map J      \HS     {\wt \HS},    & \qquad & \map  {J'} {\wt \HS} \HS\\
    \map {J_1} {\HS_1} {\wt \HS_1}, & \qquad & \map {J_1'} {\wt \HS_1} {\HS_1}.
  \end{array}
  \end{equation}
  Let $\delta>0$ and $k \ge 1$. We say that $(H,\HS)$ and $(\wt H,\wt
  \HS)$ are \emph{$\delta$-close} with respect to the
  \emph{quasi-unitary maps} $(J,J_1)$ and $(J',J_1')$ of \emph{order
    $k$} iff the following conditions are fulfilled:
\begin{align}
  \label{eq:j.scale}
  \norm[1 \to 0] {J - J_1} &\le \delta, & 
  \norm[1 \to 0] {J'-J_1'} &\le \delta\\
  \label{eq:j.adj}
  \norm[0 \to 0] {J-{J'}^*} &\le \delta, &\\
  \label{eq:j.comm}
  \norm[k \to -1] {\wt H J_1 - J_1'^* H} & \le \delta, &\\ 
  \label{eq:j.inv}
  \norm[1 \to 0] {\1 - J' J} & \le \delta, & 
  \norm[1 \to 0] {\1 - J J'} & \le \delta\\ 
  \label{eq:j.bdd}
  \norm[0 \to 0] J   & \le 2, & 
  \norm[0 \to 0] {J'} & \le 2.
\end{align}
\end{definition}

Note that all operators make sense on the given domains, e.g.,
\begin{displaymath}
   \map{(\wt H J_1 - J_1'^* H) \restr {\HS_k} = 
        (\wt H J_1 - J_1'^* H) \1_{k \to 1}} 
           {\HS_k} {\wt \HS_{-1}} 
\end{displaymath}
where $\map {\1_{k \to 1}} {\HS_k} {\HS_1}$ is the natural inclusion
map. Strictly speaking, we should also write $J \1_{1 \to 0} - \wt \1_{1
  \to 0} J_1$ in \eqref{eq:j.scale} and $(\1_{0 \to 0} - J' J)\1_{1
  \to 0}$ in \eqref{eq:j.inv}, e.g., but we refrain from it in order
to keep the notation readable.

We can interprete~\eqref{eq:j.comm} in the sense that $J_1$ and $J'_1$
are \emph{quasi-intertwining} operators.  Since~\eqref{eq:j.scale}
assures the closeness of $J$ and $J_1$ resp.\ $J'$ and $J_1'$ we call
$J$ resp.\ $J'$ also \emph{quasi-intertwining}.
Furthermore,~\eqref{eq:j.inv} together with~\eqref{eq:j.adj} says that
$J$ and $J'$ are \emph{quasi-unitary}. In our application of a
graph-like manifold converging to a metric graph, cf.\ \Sec{graph}),
the regularity order $k$ equals $1$. In this case, the assumptions are
symmetric in $H$ and $\wt H$, but we will also meet situations in a
forthcoming paper where $k>1$ is needed.

For concrete applications, the following equivalent characterisation
of \eqref{eq:j.scale}--\eqref{eq:j.bdd} will be useful:
\begin{align}
  \label{eq:j.scale.}
  \tag{\ref{eq:j.scale}'}
  \norm[0]{Jf-J_1f} &\le \delta \norm[1] f, &
  \norm[0]{J' u-J_1' u} &\le \delta \norm[1] u\\
  \label{eq:j.adj.}
  \tag{\ref{eq:j.adj}'}
  |\iprod {Jf}  u - \iprod f {J' u}| & \le
     \delta \norm[0] f \norm[0] u\\
  \label{eq:j.comm.}
  \tag{\ref{eq:j.comm}'}
  |\wt{\qf h} (J_1 f, u) - \qf h (f, J_1' u)| & \le
     \delta \norm[k] f \norm[1] u\\
  \label{eq:j.inv.}
  \tag{\ref{eq:j.inv}'}
  \norm[0] {f - J' J f} & \le \delta \norm[1] f, & 
  \norm[0] {u - J J' u} & \le \delta\ \norm[1] u\\ 
  \label{eq:j.bdd.}
  \tag{\ref{eq:j.bdd}'}
  \norm[0] {J f}   & \le 2 \norm[0] f, & 
  \norm[0] {J' u} & \le 2 \norm[0] u
\end{align}
for all $f, u$ in the appropriate spaces. Here, $\qf h$ and
$\wt{\qf h}$ denote the sesquilinear forms associated to $H$ and $\wt
H$, i.e., $\qf h(f,g)=\iprod {H^{1/2}f} {H^{1/2}g}$ for $f,g \in
\HS_1$ and similarly for $\wt{\qf h}$.

Let us illustrate the above abstract setting in the following example
of norm resolvent convergence in a fixed Hilbert space:
\begin{example}
  \label{ex:nrconv}
  Suppose that $\wt \HS = \HS$, $J=J' = \1$, $J_1 = J_1' = \1$, $k=1$
  and $\delta = \delta_n \to 0$ as $n \to \infty$. Assume in addition
  that the quadratic form domains of $H$ and $\wt H = H_n$ agree. Now
  the only non-trivial assumption in \Def{closeness} is \Eq{j.comm},
  which is equivalent to
  \begin{displaymath}
    \norm[1 \to -1]{H_n - H} = 
    \norm[0 \to 0] {R_n^{1/2}(H_n - H) R^{1/2}} \to 0
  \end{displaymath}
  whereas $H_n \to H$ in norm resolvent convergence means
  \begin{displaymath}
    \norm[0 \to 0] {R_n - R} = 
    \norm[0 \to 0] {R_n(H_n - H) R} =
    \norm[2 \to -2] {H_n - H} \to 0
  \end{displaymath}
  as $n \to \infty$.  Therefore, we see that our
  assumption~\eqref{eq:j.comm} implies the norm resolvent convergence
  but not vice versa.
\end{example}

\begin{remark}
  \label{rem:closeness}
  We have expressed the closeness of certain quantities in dependence
  on the initial closeness data $\delta>0$.  Although, in our
  applications, $(\wt H,\wt \HS)$ will depend on some parameter $\eps>0$
  with $\delta=\delta(\eps) \to 0$ as $\eps \to 0$ we prefer to
  express the dependence only in terms of $\delta$. In
  particular, an assertion like $\norm{JR - \wt R J} \le 4 \delta$
  means that it is true for all $(H,\HS)$ and $(\wt H,\wt \HS)$ being
  $\delta$-close with respect to $(J,J_1)$ and $(J',J_1')$. In this
  sense, $(H,\HS)$ and $(\wt H,\wt \HS)$ should be considered as
  ``variables'' being close to each other.
\end{remark}

We deduce the following simple estimates:
\begin{lemma}
  \label{lem:j.iso}
  Suppose that Assumption~\eqref{eq:j.adj}, \eqref{eq:j.inv}
  and~\eqref{eq:j.bdd} are fulfilled, then
  \begin{equation}
    \label{eq:j.iso}
    \norm[0] f - \delta' \norm[1] f \le 
    \norm[0]{Jf} \le 
    \norm[0] f + \delta' \norm[1] f \qquad \text{with} \qquad
    \delta' := \sqrt{3\delta}
  \end{equation}
  and similarly for $J'$.
\end{lemma}

\begin{proof}
  We calculate
  \begin{multline*}
    \big| \normsqr{Jf} - \normsqr f \big| =
    \big| \iprod {(J^*J - \1)f} f \big| \le
    \big| \iprod {(J^* - J')Jf} f \big| +
       \big| \iprod {(J'J - \1) f} f \big| \\ \le
    \norm[0 \to 0] {J^* - J'} \norm[0]{Jf} \norm[0] f +
       \norm[1 \to 0]{J' J - \1} \norm[1] f \norm[0] f \le
    3\delta \normsqr[1] f
  \end{multline*}
  and the result follows.
\end{proof}

\subsection{Resolvent convergence and functional calculus}
\label{sec:res.conv}

In this section we prove our result on resolvent convergence.  More
precisely, we estimate the errors in terms of $\delta$.  All the
results below are valid for pairs of non-negative operators and
Hilbert spaces $(H,\HS)$ and $(\wt H,\wt\HS)$ which are $\delta$-close
of order $k$. We set
\begin{equation}
  \label{eq:m}
   m := \max \{ 0, k-2\} 
\end{equation}
as regularity order for the resolvent difference. Note that $m=0$ if
$k=1$ (as in our application) or $k=2$.
\begin{theorem}
\label{thm:res}
  Suppose~\eqref{eq:j.scale}, \eqref{eq:j.adj} and~\eqref{eq:j.comm}, then
  \begin{gather}
  \label{eq:res}
    \norm[m \to 0] {\wt R J - JR}  = 
    \norm[2+m \to -2]{JH - \wt H J} \le 4\delta,\\
  \label{eq:res.j}
    \norm[m \to 0] {\wt R^j J - JR^j}  \le 4j \delta
  \end{gather}
  for all $j \in \N$.
\end{theorem}

\begin{proof}
  We start with the equation
  \begin{equation*}
    JH - \wt H J = 
    (J - J'^*)H + (J'-J_1')^*H + (J_1'^* H - \wt H J_1) + \wt H (J_1 - J)
  \end{equation*}
  considered as bounded operator from $\HS_{2+m}$ to $\wt \HS_{-2}$.
  Using~\eqref{eq:norm.adj} and~\eqref{eq:norm.op.ineq} yields
  \begin{multline*}
    \norm[m \to 0]{\wt R J - JR} =
    \norm[m \to 0]{\wt R (JH - \wt H J) R} =
    \norm[2+m \to -2] {JH - \wt H J} \\\le
    \norm[m \to -2] {J-{J'}^*} + 
        \norm[2 \to -m] {J' - J_1'} + 
        \norm[2+m \to -2] {J_1'^* H - \wt H J_1} +
        \norm[2+m \to 0] {J_1 - J} \\ \le
    \norm[0 \to 0] {J-{J'}^*} + 
        \norm[1 \to 0] {J' - J_1'} + 
        \norm[k \to -1] {J_1'^* H - \wt H J_1} +
        \norm[1 \to 0] {J_1 - J} \le
    4\delta,
  \end{multline*}
  i.e., the assertion~\eqref{eq:res}. For the second estimate we use
  the resolvent identity
  \begin{displaymath}
    \wt R^j J - J R^j = 
    \sum_{i=0}^{j-1} \wt R^{j-1-i} (\wt R J - JR) R^i,
  \end{displaymath}
  and conclude
  \begin{equation*}
    \norm[m \to 0] {\wt R^j J - JR^j} \le
    \sum_{i=0}^{j-1}
         \norm[0 \to 0] {\wt R^{j-1-i}}
         \norm[m \to 0] {\wt R J - JR} 
         \norm[m \to m] {R^i} \le
     4 j \delta
  \end{equation*}
  using the estimate for $j=1$. Note that $\norm[m \to m] R \le 1$ for
  any $m$ and similarly for $\wt R$.
\end{proof}

\begin{remark}
  Observe that we cannot obtain a better result using the
  quasi-unitary operator $J$ although we loose regularity order at
  some stages. The best what we can expect (in the case $k=1$)
  is the estimate
  \begin{equation}
    \label{eq:res.better}
     \norm[-1 \to 1]{\wt R J_1'^* - J_1R} \le 4 \delta    
  \end{equation}
  which follows from
  \begin{multline*}
    (\wt H + 1)^{1/2}(\wt R J_1'^* - J_1 R)(H + 1)^{1/2} \\=
    \wt R^{1/2} \bigl[
       (J_1'^*H - \wt H J_1) + 
          (J_1'^* - J'^*) + (J'^* - J) + (J - J_1)
    \bigr] R^{1/2}
  \end{multline*}
  and the assumptions.
  
  On the other hand, if we assume that $\norm[2+m \to -2] {J H - \wt H
    J} \le \wt \delta$, i.e.,
  \begin{equation}
    \label{eq:j.op}
    | \iprod {JHf} u - \iprod {Jf} {\wt H u}| \le 
    \wt \delta \norm[2+m] f \norm[2] u
  \end{equation}
  for all $f \in \HS_{2+m}$, $u \in \wt \HS_2$ then we directly obtain
  the resolvent estimate~\eqref{eq:res} with $\wt \delta = 4\delta$.
  Although Assumption~\eqref{eq:j.op} is weaker
  than~\eqref{eq:j.scale}--\eqref{eq:j.comm} (cf.\ \Ex{nrconv}) it is
  often easier in our applications to deal with the quadratic form
  domains, even if one needs the additional operators $J'$, $J_1$ and
  $J_1'$ and the stronger
  estimates~\eqref{eq:j.scale}--\eqref{eq:j.comm}.
\end{remark}

We want to extend our results to more general functions $\phi(H)$ of
the operator $H$ and similarly for $\wt H$. We start with continuous
functions on $\R_+ := [0, \infty)$ such that $\lim_{\lambda \to
  \infty} \phi(\lambda)$ exist, i.e., with functions continuous on
$\overline \R_+ := [0, \infty]$. We denote this space by $\Cont
{\overline \R_+}$.

\begin{theorem}
\label{thm:cont}
Suppose that~\eqref{eq:j.scale}, \eqref{eq:j.adj}, \eqref{eq:j.comm}
and~\eqref{eq:j.bdd} are fulfilled, then
  \begin{equation}
    \label{eq:cont}
    \norm[m \to 0] {\phi(\wt H) J - J \phi(H)} \le \eta_\phi(\delta)
  \end{equation}
  for all $\phi \in \Cont {\overline \R_+}$ where $\eta_\phi(\delta)
  \to 0$ as $\delta \to 0$.
\end{theorem}

\begin{proof}
  Let $p(\lambda) := \sum_{j=0}^n a_j (\lambda+1)^{-j}$ be a
  polynomial in $(\lambda+1)^{-1}$. Then
  \begin{multline*}
    \norm[m \to 0]{\phi(\wt H) J - J\phi(H)} \\ \le
    \norm[0 \to 0]{(\phi - p)(\wt H)} \norm[m \to 0] J +
       \norm[0 \to 0] J \norm[m \to 0] {(\phi - p)(H)} \qquad\qquad \\ {}+
       \sum_{j=0}^n |a_j| \norm[m \to 0] {\wt R^j J - JR^j} \le
    4 \norm[\infty] {\phi - p} + \sum_{j=0}^n |a_j| 4 j \delta =:
   \eta_\phi(\delta,p)
  \end{multline*}
  using~\eqref{eq:norm.op.ineq}, the spectral calculus,
  \eqref{eq:j.bdd} and~\eqref{eq:res.j}. Here, $\norm[\infty] \phi$
  denotes the supremum norm of $\phi$.
  
  Suppose $\eta > 0$. By the Stone-Weierstrass theorem there exists a
  polynomial $p$ such that $\norm[\infty]{p - \phi} \le \eta/8$. If
  \begin{displaymath}
    0 < \delta \le \frac \eta {8 \sum_{j=0}^n |a_j| j}    
  \end{displaymath}
  then $\eta_\phi(\delta) := \eta_\phi(\delta,p) \le \eta/2 + \eta/2
  = \eta$ and therefore $\eta_\phi(\delta) \to 0$ as $\delta \to 0$.
\end{proof}

In a second step we extend the previous result to certain bounded
measurable functions $\map \psi {\overline \R_+} \C$.
\begin{theorem}
  \label{thm:meas}  
  Suppose that $U \subset \overline \R_+$ and that $\map \psi
  {\overline \R_+} \C$ is a measurable, bounded function, continuous
  on $U$ such that $\lim_{\lambda \to \infty} \psi(\lambda)$ exist.
  Then
  \begin{equation}
    \label{eq:meas}
    \norm[m \to 0] {\psi(\wt H) J - J \psi(H)}  \le \eta_\psi(\delta)
  \end{equation}
  for all pairs of non-negative operators and Hilbert spaces $(H,\HS)$
  and $(\wt H, \wt \HS)$ which are $\delta$-close provided
  \begin{equation*}
    \spec H \subset U \qquad \text{or} \qquad \spec {\wt H} \subset U.
  \end{equation*}
  Furthermore, $\eta_\phi(\delta) \to 0$ as $\delta \to 0$.
\end{theorem}

\begin{proof}
  Let $\chi_1$ be a continuous function on $\overline \R_+$ satisfying
  $0 \le \chi_1 \le 1$, $\chi_1 = 1$ on $\spec H \cup \{\infty\}$
  (resp.\ $\chi_1 = 1$ on $\spec {\wt H} \cup \{\infty\}$ if $U$ is a
  neighbourhood of $\spec {\wt H}$) and $\supp \chi_1 \subset U$. Then
  $\chi_1 \psi$ and $\chi_2 = 1 - \chi_1$ are continuous functions on
  $\overline \R_+$ and
  \begin{multline*}
    \norm[m \to 0]{\psi(\wt H) J - J\psi(H)} \\ \le
    \norm[m \to 0]{(\chi_1\psi)(\wt H)J - J (\chi_1\psi)(H)} +
       \norm[m \to 0]{(\chi_2\psi)(\wt H)J - J (\chi_2\psi)(H)}.
  \end{multline*}
  In the case that $U$ is a neighbourhood of $\spec H$ we can estimate
  the norm with $\chi_2$ by
  \begin{displaymath}
   \norm[\infty]{\psi} 
      \norm[m \to 0]{\chi_2(\wt H)J - J \chi_2(H)}
  \end{displaymath}
  using the fact that $(\chi_2 \psi)(H) = \chi_2(H) = 0$ since
  $\chi_2=0$ on $\spec H$.
  
  In the case that $U$ is a neighbourhood of $\spec {\wt H}$ and if $m \ge
  1$ then we can estimate the norm with $\chi_2$ by
  \begin{displaymath}
      \norm[m \to 0]{J(\chi_2 \psi)(H)} \le
         \norm[0 \to 0] J \norm[m \to 0]{(\chi_2 \psi)(H)} \le
         2 \norm[\infty] \psi \norm[m \to 0]{\chi_2(H)}
  \end{displaymath}
  again using the fact that $(\chi_2 \psi)(\wt H) = \chi_2(\wt H) = 0$
  since $\chi_2=0$ on $\spec {\wt H}$. Now
  \begin{equation*}
    \norm[m \to 0]{\chi_2(H)} \le
    \norm[1 \to 0] {\1 - J' J} \norm[m \to 1] {\chi_2(H)} + {}
      \norm[0 \to 0] {J'} \norm[m \to 0] {J \chi_2(H) - \chi_2(\wt H) J}.
  \end{equation*}
  Note that $\norm[m \to 1] {\chi_2(H)} \le 1$ since $m \ge 1$. If
  $m=0$ then use the fact that
  \begin{displaymath}
    \norm[0 \to 0]{\psi(\wt H) J - J\psi(H)} =
    \norm[0 \to 0]{\psi(H) J^* - J^* \psi(\wt H)}
  \end{displaymath}
  and argue as in the case where $\spec H \subset U$ with the roles of
  $H$ and $\wt H$ interchanged.

  Applying the preceding 
  theorem twice (in each of the above cases), we
  have the error estimate
  \begin{displaymath}
    \eta_\psi(\delta):=
      \eta_{\chi_1 \psi}(\delta) + 
        2 \norm[\infty]{\psi} (2 \eta_{\chi_2}(\delta) + \delta).
  \end{displaymath}
\end{proof}

\begin{example}
  Consider $\psi=\1_I$ with an interval $I$ such that $\bd I \cap
  \spec H = \emptyset$ or $\bd I \cap \spec {\wt H} = \emptyset$ then
  the spectral projections satisfy
  \begin{equation}
    \label{eq:spec.proj}
    \norm[m \to 0]{\1_I(\wt H) J - J \1_I(H)} \le \eta_{\1_I}(\delta).
  \end{equation}
\end{example}

Finally we show the following estimates from the ones already
considered:
\begin{theorem}
  \label{thm:other.est}
  Suppose that \eqref{eq:j.adj}, \eqref{eq:j.inv}, \eqref{eq:j.bdd}
  and
  \begin{displaymath}
    \norm[m \to 0] {\phi(\wt H) J - J \phi(H)} \le \eta
  \end{displaymath}
  for some function $\phi$ and some constant $\eta>0$. Then we have
  \begin{align}
    \label{eq:dual}
    \norm[0 \to -m] {\phi(H) J' - J' \phi(\wt H)} & \le 
        2\norm[\infty] \phi \delta + \eta\\
    \label{eq:sandwich}
    \norm[m \to 0] {\phi(H) - J' \phi(\wt H) J} & \le 
        C \delta + 2 \eta\\
    \label{eq:sandwich.}
    \norm[0 \to 0] {\phi(\wt H) - J \phi(H) J'} & \le 
        5C \delta + 2 \eta
  \end{align}
  provided $m=0$ for the last estimate. Here, $C:= \norm[\infty]
  \phi$ if $m\ge 1$ and $C>0$ is a constant satisfying
  $|\phi(\lambda)| \le C (\lambda+1)^{-1/2}$ for all $\lambda$ if
  $m=0$.
\end{theorem}

\begin{proof}
  The first estimate follows from
  \begin{displaymath}
    \norm[0 \to -m] {\phi(H) J' - J' \phi(\wt H)} \le
    2 \norm[\infty] \phi \norm[0 \to 0] {J' - J^*} +
      \norm[0 \to -m] {\phi(H) J^* - J^* \phi(\wt H)}
  \end{displaymath}
  and~\eqref{eq:norm.adj}; the second from
  \begin{multline*}
    \norm[m \to 0] {\phi(H) - J' \phi(\wt H) J} \\ \le
    \norm[1 \to 0] {\1 - J' J} \norm[m \to 1] {\phi(H)}
      + \norm[0 \to 0] {J'} \norm[m \to 0] {J \phi(H) - \phi(\wt H) J}
  \end{multline*}
  and the third from
  \begin{multline*}
    \norm[0 \to 0] {\phi(\wt H) - J \phi(H) J'} \\ \le
    \norm[1 \to 0] {\1 - J J'} \norm[0 \to 1] {\phi(\wt H)}
      + \norm[0 \to 0] J \norm[0 \to 0] {J' \phi(\wt H) - \phi(H) J'}
  \end{multline*}
  together with~\eqref{eq:dual}.
\end{proof}

\subsection{Spectral convergence}
\label{sec:spec}

We now prove some convergence results for spectral projections and
(parts) of the spectrum.
\begin{theorem}
  \label{thm:proj}
  Let $I$ be a measurable and bounded subset of $\R$.  Then there
  exists $\delta_0=\delta_0(I,k)>0$ such that for all $\delta>0$ we
  have
  \begin{equation*}
    \dim P = \dim \wt P
  \end{equation*}
  for all pairs of non-negative operators and Hilbert spaces $(H,\HS)$
  and $(\wt H,\wt \HS)$ which are $\delta$-close of order $k$ provided
  \begin{equation*}
    \bd I \cap \spec H = \emptyset \qquad \text{or} \qquad 
    \bd I \cap \spec {\wt H}  = \emptyset.
  \end{equation*}
  Here, $P := \1_I(H)$ and $\dim P := \dim P(\HS)$, similarly for
  $\wt H$.
\end{theorem}
\begin{proof}
  Let us first show the inequality $\dim P \le \dim \wt P$:
  Suppose $f \in P(\HS)$. Then $\norm[m] f \le C_{I,m} \norm[0] f$ with
  \begin{displaymath}
    C_{I,m} := \sup_{\lambda \in I} (1+\lambda)^{m/2} < \infty
  \end{displaymath}
  since $I$ is bounded. Furthermore,
  \begin{multline*}
    \norm[0] {\wt P J f} \ge 
    \norm[0] {J P f} - \norm[0] {(\wt P J - JP)f} \\ \ge
    \norm[0] {J f} - \norm[m \to 0] {\wt P J - J P} \norm[0] f \ge
    (1 - \delta' C_{I,1} - \eta_{\1_I}(\delta)) \norm[0] f
  \end{multline*}
  using \Lem{j.iso} and \Thm{meas}. Since $\delta' \to
  0$ and $\eta_{\1_I}(\delta) \to 0$ as $\delta \to 0$ there exists
  $\delta_0 > 0$ such that
  \begin{equation}
    \label{eq:pj.lower}
    \norm[0] {\wt P J f} \ge \frac 12 \norm[0] f
  \end{equation}
  provided $0 < \delta \le \delta_0$. Therefore, $\wt P J \restr
  {P(\HS)}$ is injective. If $f_1, \dots, f_d$ are linear independent in
  $P(\HS)$, the same is true for $\wt P J f_1, \dots, \wt P J f_d$ in
  $\wt P(\wt \HS)$. If $P(\HS)$ is infinite dimensional so is
  $\wt P(\wt \HS)$. Thus we have shown $\dim P \le \dim \wt P$.
  
  The other inequality is more difficult due to the asymmetry in the
  norm convergence $\norm[m \to 0] \cdot$ if $m>0$. Suppose that $u
  \in \wt P(\wt \HS)$ and that $\chi_i \in \Cont {\overline \R_+}$ with
  $\chi_1 + \chi_2 + \chi_3 = 1$. Suppose in addition that $\supp
  \chi_1$ and $\supp \chi_2$ are compact, that $\supp \chi_1$ and
  $\supp \chi_3$ are disjoint and that $\supp \chi_2 \cap I =
  \emptyset$. Then
  \begin{multline*}
    \norm[-m] {P J^* u} \ge 
    \norm[-m] {J^* \wt P u} - \norm[-m] {(\wt P J^* - J^* \wt P)u} \\ \ge
    \norm[-m] {\chi_1(H) J^* u} - 
        \norm[-m] {\chi_2(H) J^* u} -
        \norm[-m] {\chi_3(H) J^* u} -
        \norm[m \to 0] {\wt P J - J P} \norm[0] u \\ \ge
     C'_{I,m} \norm[0] {J^* \wt P u} -
        \norm[-m] {(\chi_2(H) J^* - J^* \chi_2(\wt H))\wt P u} -
        \eta_{\1_I}(\delta) \norm[0] u
  \end{multline*}
  by \Thm{meas} and the fact that $\chi_2(\wt H) \wt P = (\chi_2
  \1_I)(\wt H) = 0$ since the support of $\chi_2$ and $I$ are
  disjoint. Here,
  \begin{displaymath}
    C'_{I,m} := \inf_{\set{\lambda}{\chi_1(\lambda)=1}} 
                            (1+\lambda)^{-m/2} - 
                \sup_{\lambda \in \supp \chi_3} (1+\lambda)^{m/2}
  \end{displaymath}
  by the spectral calculus. Since $\chi_1$ and $\chi_3$ have disjoint
  support, $C'_{I,m}>0$. Next, the norm involving $\chi_2$ can be
  estimated from above by
  \begin{displaymath}
    \eta_{\chi_2}(\delta) \norm[0] u
  \end{displaymath}
  using \Thm{cont}. Furthermore,
  \begin{displaymath}
    \norm[0] {J^* u} \ge 
    \norm[0] {J' u} - \norm[0] {(J^* - J') u} \ge
    (1 - C_{I,1} \delta' - \delta) \norm[0] u
  \end{displaymath}
  by \Lem{j.iso} and \eqref{eq:j.adj}. Finally, we have shown that
  \begin{displaymath}
    \norm[-m] {P J^* u} \ge 
    \big(
       C'_{I,m}(1 - C_{I,1} \delta' - \delta) - \eta_{\chi_2}(\delta) -
          \eta_{\1_I} (\delta)
    \big)
    \norm[0] u.
  \end{displaymath}
  The inequality $\dim P \ge \dim \wt P$ follows as before.
\end{proof}

In the case of $1$-dimensional projections we can even show the
convergence of the corresponding eigenvectors. Note that generically,
the eigenvalues are simple (cf.~\cite{uhlenbeck:76}):
\begin{theorem}
  \label{thm:eigenvectors}
  Suppose that $\phi$ is a normalised eigenvector of $H$ with
  eigenvalue $\lambda$ and that $\dim \1_I(H)=1$ for some open,
  bounded interval $I \subset [0,\infty)$ containing $\lambda$. Then
  there exists $\delta_0=\delta(I,k)>0$ such that $\wt H$ has only one
  eigenvalue $\wt \lambda$ of multiplicity $1$ in $I$ for all $(\wt H,
  \wt \HS)$ being $\delta$-close of order $k$ to $(H,\HS)$ and all
  $0<\delta<\delta_0$.
  
  In addition, there exist a unique eigenvector $\wt \phi$ (up to a
  unitary scalar factor close to $1$) and functions
  $\eta_{1,2}(\delta) \to 0$ as $\delta \to 0$ depending only on
  $\lambda$ and $k$ such that
  \begin{align*}
    \norm{J \phi - \wt \phi} &\le \eta_1(\delta), &
    \norm{J' \wt \phi - \phi} &\le \eta_2(\delta).
  \end{align*}
\end{theorem}
\begin{proof}
  Denote the corresponding eigenprojections by $P$ resp.\ $\wt P$. The
  first assertion follows from \Thm{proj}. For the second, note that
  \begin{equation*}
    \wt \phi = \frac 1 {\iprod {\wt P J \phi} {J\phi}} \, \wt P J \phi
  \end{equation*}
  since $\wt P$ is a $1$-dimensional projection. Note in addition that
  \begin{equation*}
    \iprod {\wt P J \phi} {J \phi} =
    \normsqr {\wt P J \phi} \ge
    \frac 1 4 \normsqr \phi =
    \frac 1 4, \qquad 0 < \delta < \delta_0
  \end{equation*}
  for some $\delta_0=\delta_0(I,k)$ due to~\eqref{eq:pj.lower}. Now,
  \begin{multline*}
    \norm{J \phi - \wt \phi} =
    \Bignorm{J P \phi - 
          \frac 1 {\iprod {\wt P J \phi} {J \phi}} \, \wt P J \phi} \\ \le
    \norm{(JP - \wt PJ)\phi} + 
       \Bigl| 1 - \frac 1  {\iprod {\wt P J \phi} {J \phi}} \Bigr|
       \norm{\wt P J \phi} \\ \le
    \eta_{\1_I}(\delta) +
        8 \bigr| \iprod {(\wt P J - J P)\phi} {J\phi} + 
           \normsqr{J \phi} - \normsqr \phi \bigr| \le
    17 \eta_{\1_I}(\delta) + 3 \delta =: \eta_1(\delta)
  \end{multline*}
  since $\phi=P\phi$ and $\norm \phi=1$ using~\eqref{eq:j.bdd}
  and~\eqref{eq:j.iso}. The second estimate follows immediately from
  \begin{equation*}
    \norm{J' \wt \phi - \phi} \le
    \norm{J'(\wt \phi - J \phi)} + \norm{(J'J-\1)\phi} \le
    2 \eta_1(\delta) + \delta(1+\lambda) =: \eta_2(\delta).
  \end{equation*}
  All estimates are valid for $0<\delta<\delta_0$. Note that
  $\delta_0$ and $\eta_i(\delta)$ depend also on $I$ and therefore on
  $\lambda$.
\end{proof}

We now show that the spectrum of the resolvents $R=(H+1)^{-1}$ and
$\wt R =(\wt H + 1)^{-1}$ are close in the Hausdorff distance defined
by
\begin{equation}
  \label{eq:hausdorff}
  d(A,B) := \max \bigl\{ \sup_{a \in A} d(a,B), \sup_{b \in B} d(b,A) \bigr\}
\end{equation}
for subsets $A,B$ of $\R$ where $d(a,B) := \inf_{b \in B} |a-b|$.
Furthermore, we set\footnote{Strictly speaking, $(\spec H + 1)^{-1} =
  \spec R \setminus \{0\}$, but the point $0$ plays no special role
  since $d(A,B)=d(\clo A,\clo B)$.}
\begin{equation}
  \label{eq:hausdorff.res}
  \overline d(A,B) := d((A+1)^{-1},(B+1)^{-1})
\end{equation}
for closed subsets of $[0, \infty)$ (cf.
also~\cite[Appendix~A]{herbst-nakamura:99}, where an equivalent
characterisation of the convergence $\overline d(A_n, A) \to 0$ as $n
\to \infty$ is given).
\begin{theorem}
  \label{thm:spectrum}
  There exists $\eta(\delta)>0$ with $\eta(\delta) \to 0$ as $\delta
  \to 0$ such that
  \begin{displaymath}
    \overline d\bigl(\spec[\bullet] H, \spec[\bullet] {\wt H}\bigr) 
           \le \eta(\delta)
  \end{displaymath}
  for all pairs of non-negative operators and Hilbert spaces $(H,\HS)$
  and $(\wt H,\wt \HS)$ which are $\delta$-close. Here, $\spec[\bullet]
  H$ denotes either the entire spectrum, the essential or the discrete
  spectrum of $H$.

  Furthermore, the multiplicity of the discrete spectrum is preserved,
  i.e., if $\lambda \in \disspec H$ has multiplicity $m>0$ then $\dim
  \1_I(\wt \HS) = m$ for $I:=(\lambda - \eta(\delta), \lambda +
  \eta(\delta))$ provided $\delta$ is small enough.
\end{theorem}
\begin{proof}
  We start with the discrete spectrum. Let $\eta>0$ and $z=(\lambda +
  1)^{-1}>0$, $\lambda \in \disspec H$. By the definition of the
  discrete spectrum, there exists an open interval $I$ containing
  $\lambda$ such that $I \cap \spec H = \{\lambda\}$ and $0 < \dim
  \1_I(H) < \infty$. Without loss of generality, we assume that $I
  \subset (\lambda - \eta, \lambda + \eta)$. From \Thm{proj} it
  follows that $\dim \1_I(H) = \dim \1_I(\wt H)$ provided $0 < \delta <
  \delta_z$ for some $\delta_z >0$. In particular, the multiplicity is
  preserved and there exists $\wt \lambda \in I \cap \disspec {\wt H}$,
  i.e.,
  \begin{equation}
    \label{eq:sp.est}
    d(z, \wt S) \le |z - \wt z| \le |\lambda - \wt \lambda| < \eta
  \end{equation}
  where $\wt S = (\disspec {\wt H} + 1)^{-1}$ and $\wt z = (\wt
  \lambda + 1)^{-1}$.  Now let $\delta(\eta)$ be the minimum of all
  $\delta_z$ where $z$ runs through the finite set $S \cap [\eta, 1]$
  with $S:=(\disspec H + 1)^{-1}$.  Then~\eqref{eq:sp.est} holds for
  all $z \in S \cap [\eta, 1]$ and $0 < \delta < \delta(\eta)$. If
  $\disspec H$ is finite, we just have to assure that $\eta < \inf S$.
  If $\disspec H$ is infinite, so is $\disspec {\wt H}$ and in
  particular, $\wt S \cap (0,\eta) \ne \emptyset$ for all $\eta>0$.
  Therefore, if $z \in (0,\eta) \cap S$ then $d(z, \wt S) \le \eta$.
  Finally,~\eqref{eq:sp.est} holds for all $z \in S$ and $0<\delta <
  \delta(\eta)$.
  
  Interchanging the roles of $H$ and $\wt H$ leads to the inequality
  $d(\wt z, S) \le \eta$ for all $\wt z \in \wt S$ and therefore
  $\overline d(\disspec H, \disspec {\wt H}) \le \eta(\delta)$ where
  $\eta(\delta)$ is the smallest constant satisfying the previous
  estimate for all $(H,\HS)$, $(\wt H, \wt \HS)$ being $\delta$-close.

  For the essential spectrum we argue similarly: Let $\eta>0$ and $z
  =(\lambda + 1)^{-1}>0$ with $\lambda \in \essspec H$. Let $I$ be an
  open interval with $\lambda \in I$ and $\bd I \cap \spec H =
  \emptyset$. If $I$ can be chosen in such a way that $I \subset
  (\lambda - \eta, \lambda + \eta)$ then $\infty = \dim \1_I(H) = \dim
  \1_I(\wt H)$ for all pairs $(\wt H, \wt \HS)$ being $\delta$-close,
  $0 < \delta < \delta_z$ for some fixed $\delta_z>0$ due to
  \Thm{proj}. In particular, $I \cap \essspec {\wt H} \ne \emptyset$
  and therefore $d(z, \wt S) < \eta$ as in~\eqref{eq:sp.est} where now
  $\wt S=(\essspec {\wt H} + 1)^{-1}$.
  
  If no such interval $I$ exist, then there is $0<\eta_0<\eta$ such
  that $I_0:=(\lambda-\eta_0, \lambda+\eta_0) \subset \essspec H$. We
  want to show that in this case, $I_0 \subset \essspec {\wt H}$ and
  in particular, $d(z, \wt S) \le \eta_0 < \eta$ provided
  $0<\delta<\delta_z$ for some fixed $\delta_z$: Suppose that this is
  not true.  Then there were $\wt \lambda \in I_0$ and an open
  interval $J$ containing $\wt \lambda$ which is disjoint from the
  closed set $\essspec {\wt H}$ for all $\delta>0$ and all $(\wt H,\wt \HS)$
  being $\delta$-close. But then, \Thm{proj} implies $0 = \dim \1_J
  (\wt H) = \dim \1_J (H)$ contradicting the fact that $J \subset
  \essspec H$.
  
  A compactness argument shows that there exists $\delta(\eta)>0$ such
  that~\eqref{eq:sp.est} is true for all $z$ in the compact set $S
  \cap [\eta, 1]$ and all $(\wt H,\wt \HS)$ being $\delta$-close,
  $\delta < \delta(\eta)$ where $S=(\essspec H + 1)^{-1}$.  If
  $\essspec H$ is bounded (from above) then $S \cap [\eta, 1] = S$
  provided $\eta < \inf S$ and we are done. If $\essspec H$ is
  unbounded, a similar reasoning as before shows that the same is true
  for $\essspec {\wt H}$. In particular, $(0, \eta) \cap \wt S \ne
  \emptyset$ and $d(z, \wt S) < \eta$ for $z \in (0,\eta) \cap S$,
  i.e.,~\eqref{eq:sp.est} holds for all $z \in S$. The assertion
  follows as in the discrete case by symmetry.
  
  The case of the entire spectrum can be shown similarly.
\end{proof}
We have the following immediate consequences when $\disspec H =
\emptyset$ resp.\ $\essspec H= \emptyset$:
\begin{corollary}
  \label{cor:ess.sp}
  Suppose that $H$ has purely essential spectrum. Then for each
  $\lambda \in \essspec H$ there is essential spectrum close to
  $\lambda$ for $\wt H$ being $\delta$-close to $H$. Either $\wt H$
  has no discrete spectrum or the discrete spectrum merges into the
  essential spectrum as $\delta \to 0$.
\end{corollary}
\begin{corollary}
\label{cor:ev.conv}
Suppose that $H$ has purely discrete spectrum denoted by $\lambda_k$
(repeated according to multiplicity). Then the infimum of the
essential spectrum of $\wt H$ tends to infinity (if there where any)
and there exists $\eta_k(\delta)>0$ with $\eta_k(\delta) \to 0$ as
$\delta \to 0$ such that
  \begin{equation}
  \label{eq:ev.est.k}
    |\lambda_k - \wt \lambda_k| \le \eta_k(\delta)
  \end{equation}
  for all $(\wt H, \wt \HS)$ being $\delta$-close. Here, $\wt
  \lambda_k$ denotes the discrete spectrum of $\wt H$ (below the
  essential spectrum) repeated according to multiplicity.
\end{corollary}
Note that the convergence $\eta_k(\delta) \to 0$ is \emph{not} uniform
in $k$. The convergence of the eigenvalues can also be seen by a
direct argument using the min-max principle:

\begin{remark}
  \label{rem:ew.better}
  If we assume that
  \begin{align}
    \label{eq:j1.quad}
    \qf h(f)       &\ge \wt {\qf h} (J_1 f) - \delta \normsqr[1] f, &
    \wt {\qf h}(u) &\ge \qf h (J_1'u) - \delta \normsqr[1] u,\\
    \label{eq:j1.norm}
    \normsqr f &\ge \normsqr {J_1 f} + \delta \normsqr[1] f, &
    \normsqr u &\ge \normsqr {J_1' u} + \delta \normsqr[1] u
  \end{align}
  we obtain the more concrete eigenvalue estimate
  \begin{equation*}
    | \lambda_k - \wt \lambda_k| \le
       \frac {\bigl(\lambda_k + 2 + 
         \frac{(\lambda_k + 2)^2}{1 - \delta(\lambda_k+1)}\delta\bigr)^2}
              {1 - \delta\bigl(\lambda_k + 1 + 
        \frac{(\lambda_k+2)^2}{1 - \delta(\lambda_k+1)}\delta\bigr)}
                \cdot \delta
                = O(\delta)
  \end{equation*}
  using the min-max principle where $O(\delta)$ depends on $\lambda_k$
  (cf.~\cite[Lemma~2.1]{exner-post:05}). Note that the
  assumptions~\eqref{eq:j1.quad} and~\eqref{eq:j1.norm} are equivalent
  to the estimates
  \begin{align}
    \label{eq:j1.quad.}
    \tag{\ref{eq:j1.quad}'}
    H - J_1^* \wt H J_1 + \delta (H+1) &\ge 0, &
    \wt H - J_1'^* H J_1' + \delta (\wt H+1) &\ge 0,\\
    \label{eq:j1.norm.}
    \tag{\ref{eq:j1.norm}'}
    J_1^* J_1 - \1 + \delta(H+1) &\ge 0, &
    J_1'^* J_1' - \1 + \delta(\wt H+1) &\ge 0
  \end{align}
  in the sense that $\map A{\HS_1}{\HS_{-1}} \ge 0$ iff $\iprod {Af} f
  \ge 0$ for all $f \in \HS_1$ and similarly on $\wt \HS$. Note
  that~\eqref{eq:j1.quad} and~\eqref{eq:j1.norm} do not follow from
  the closeness assumptions~\eqref{eq:j.scale}--\eqref{eq:j.bdd};
  e.g.\ for~\eqref{eq:j1.norm} one needs in addition that $\norm[1 \to
  1]{J_1} \le C$ for some constant $C>0$ and similarly for $J_1'$. The
  estimates~\eqref{eq:j1.quad}--\eqref{eq:j1.norm} have been used
  e.g.\ in~\cite{exner-post:05, kuchment-zeng:01,
    rubinstein-schatzman:01} in the graph model and the verification
  of~\eqref{eq:j1.quad}--\eqref{eq:j1.norm} is quite similar to the
  proof of the closeness
  assumptions~\eqref{eq:j.scale}--\eqref{eq:j.bdd} as we have seen
  in~\Sec{graph}.
\end{remark}

\subsection*{Acknowledgments}
It is a pleasure to thank Pavel Exner, Peter Hislop, Fernando Lled\'o,
Gianfausto Dell'Antonio and Luca Tenuta for fruitful discussions and
helpful comments on this manuscript. Part of this work has been done
while the author was funded by the DFG under the grant Po-1034/1-1.


\providecommand{\bysame}{\leavevmode\hbox to3em{\hrulefill}\thinspace}
\providecommand{\MRhref}[2]{%
}
\providecommand{\href}[2]{#2}

\end{document}